\documentclass[acmsmall]{acmart}

\usepackage{algorithmic}
\usepackage{graphicx}
\usepackage{textcomp}
\usepackage{xcolor}
\usepackage{booktabs} 
\usepackage{multirow} 
\usepackage{xspace}
\usepackage{balance}
\usepackage{url}
\usepackage{subcaption}
\usepackage{multirow}
\usepackage[font=footnotesize]{subcaption}
\usepackage[ruled,linesnumbered]{algorithm2e}
\usepackage[capitalize,noabbrev]{cleveref}
\usepackage[normalem]{ulem}
\usepackage{colortbl}
\usepackage{xcolor}
\usepackage{framed}

\newcommand{\revise}[1]{\textcolor{black}{#1}}
\newcommand{\checkme}[1]{\textcolor{black}{#1}}
\newcommand{\checknumber}[1]{\textcolor{black}{#1}}

\newenvironment{result}{\begin{framed}\centering\it}{\end{framed}}

\AtBeginDocument{%
  }

\acmPrice{15.00}
\acmISBN{978-1-4503-XXXX-X/18/06}

\begin{document}

%\title{Latent Mutants: A large-scale study on the Interplay between mutation testing and software evolution              \\}
\title[Latent Mutants]{Latent Mutants: A large-scale study on the Interplay between mutation testing and software evolution}

\author{Jeongju Sohn}
\email{jeongju.sohn@knu.ac.kr}
\affiliation{%
  \institution{Kyungpook National University}
  \country{Republic of Korea}
}

\author{Ezekiel Soremekun}
\email{ezekiel.soremekun@rhul.ac.uk}
\affiliation{%
  \institution{Royal Holloway, University of London}
  %Royal Holloway, University of London (RHUL), UK.
  \country{UK}
}

\author{Michail Papadakis}
\email{michail.papadakis@uni.lu}
%\orcid{https://orcid.org/0000-0003-1852-2547}
\affiliation{%
  \institution{SnT, University of Luxembourg}
  \country{Luxembourg}
}

\begin{abstract}
In this paper we apply mutation testing in an in-time fashion, i.e., across multiple project releases. Thus, we investigate how the mutants of the current version behave in the future versions of the programs. We study the characteristics of what we call \textit{latent} mutants, i.e., the mutants that are live in one version and killed in later revisions, and explore whether they are predictable with these properties.  
%\fixme{The final goal is to generate a dataset that can be used to simulate hidden faults in software that may incur future failures.}{rephrase to drop simulation?} 
%Hence, our interest mainly lies in the characteristics of the mutants that survive but are revealed in later versions and whether we can identify them in advance to improve the efficiency of mutation testing. 
We examine 131,308 mutants generated by Pitest on 13 open-source projects. Around 11.2\% of these mutants are live, and 3.5\% of them are latent, manifesting in 104 days on average. Using the mutation operators and change-related features we successfully demonstrate that these latent mutants are identifiable, predicting them with an accuracy of 86\% and a balanced accuracy of 67\% using a simple random forest classifier.
%We further investigate the usability of the collected mutants through the case study on identifying bug-inducing commits using the state-of-the-art approach. The results show that while the accuracy of the approach degrades in these mutants, the general findings remain the same. 
\end{abstract}

%%
%% The code below is generated by the tool at http://dl.acm.org/ccs.cfm.
%% Please copy and paste the code instead of the example below.Latexmk: Sometimes, the -f option can be used to get latexmk
%%
%\begin{CCSXML}
%<ccs2012>
% <concept>
%  <concept_id>00000000.0000000.0000000</concept_id>
%  <concept_desc>Do Not Use This Code, Generate the Correct Terms for Your Paper</concept_desc>
%  <concept_significance>500</concept_significance>
% </concept>
% <concept>
%  <concept_id>00000000.00000000.00000000</concept_id>
%  <concept_desc>Do Not Use This Code, Generate the Correct Terms for Your Paper</concept_desc>◊
%  <concept_significance>300</concept_significance>
% </concept>
% <concept>
%  <concept_id>00000000.00000000.00000000</concept_id>
%  <concept_desc>Do Not Use This Code, Generate the Correct Terms for Your Paper</concept_desc>
%  <concept_significance>100</concept_significance>
% </concept>
% <concept>
%  <concept_id>00000000.00000000.00000000</concept_id>
%  <concept_desc>Do Not Use This Code, Generate the Correct Terms for Your Paper</concept_desc>
%  <concept_significance>100</concept_significance>
% </concept>
%</ccs2012>
%\end{CCSXML}
%
%\ccsdesc[500]{Do Not Use This Code~Generate the Correct Terms for Your Paper}
%\ccsdesc[300]{Do Not Use This Code~Generate the Correct Terms for Your Paper}
%\ccsdesc{Do Not Use This Code~Generate the Correct Terms for Your Paper}
%\ccsdesc[100]{Do Not Use This Code~Generate the Correct Terms for Your Paper}

%%
%% Keywords. The author(s) should pick words that accurately describe
%% the work being presented. Separate the keywords with commas.
%\keywords{XXX}
\maketitle
%\received{20 February 2007}
%\received[revised]{12 March 2009}
%\received[accepted]{5 June 2009}

%\todo{check all table and figure captions are correct, and all figures and tables are references in the text}

\section{Introduction} % (fold)
\label{sec:introduction}

Mutation testing is a popular test adequacy criterion that is frequently considered as one of the most effective criteria available today \cite{PapadakisK00TH19, 0020331}. A mutant is a version of the program under analysis that has been modified based on predefined syntactic transformations called mutant operators. These transformations aim at deviating the functionality of the program under analysis and thus defining the specific parts of the program functionality that should be checked.

A mutant is termed killed when a test suite can differentiate its behavior/functionality from that of the original program. Similarly, a mutant is termed live when a test suite cannot differentiate its behavior. The idea is that developers are going to inspect the live mutants and write tests that kill them. This means that mutants offer concrete guidance on what is worth testing and thus, should be targeted by test writing \cite{0020331, ChekamPTH17}.

While mutation testing can be effective, at its core lies the problem of the infinite possible mutant instances that can be used. The number of mutant instances is infinite since the application of additive operators (transformations that add code) can transform any program to any program. With that in mind we can imagine that almost every possible test case can eventually couple with a mutant instance thereby leading to a conclusion that every possible test is important (worth writing a test for) since every test kills some mutants.  

Even if the above is not true for every possible test, it is certainly true for a very large number of tests, all of which would be consider important if we consider that every test uniquely killing mutants is important, as implied by mutation testing theory. This raises the question of which mutants add the most value to the developers (in relation to test generation) under their development process constraints. This is in line with actual experience that has shown that not all mutants are perceived by developers as being useful/productive \cite{PetrovicI18}.  In other words, there are many mutants for which developers do not see much of value in writing tests for. 

In view of this, the above question can be formulated as to which mutants meet the expectations of the developers, in a sense which mutants can elicit tests that developers would spent time writing. We answer this question by introducing the notion of latent mutants, which is the set of mutants that is surviving the current test suite but not a future one. We argue that since latent mutants are killed by tests that were added later in the project timeline, i.e., latent mutants couple with future tests, developers see some value in writing these tests and thus, they are more valuable to them than the other mutants. 

A key advantage of latent mutants is that by definition they can elicit effective tests (from the developers point of view) while being relatively few (keeping the mutation testing overheads low) and with minimal semantic overlaps in the functionality deviations they incur. An additional advantage of latent mutants is that they couple with real latent faults, i.e., faults that are introduced and found at different points in the projects timeline. If we assume, as suggested by previous work \cite{PapadakisSYB18, LaurentGV22, JustJIEHF14}, that mutants couple with real faults, then by definition latent mutant will also couple with real faults since they would be live in the faulty program version and killed by a test case that is added later in the project timeline to witness the faults. %\checkme{We provide empirical evidence that for 207 (43.8\%) out of 473 faults in Defects4J~\cite{Just:2014aa} that we studied for the propagation to future versions, 115 of them (55.5\%) have such coupled latent mutants.}\fixme{}{include in the discussion?}%\footnote{We use the default group of Pitest and mutate only the fixed files in each bug to have a focused study.} 

Another interesting aspect of latent mutants is that they co-evolve together with the evolution of the software. In some sense the study of latent mutants is the study of the software evolution and its mutants for a large part of the projects' life-cycle. To this end, we apply mutation testing \textit{in time} (across different commits or program versions) by generating mutants at a particular point in time and then co-evolve them alongside the software changes. Naturally, such an ``in-time'' analysis leads to a lifespan characteristic of mutants, i.e., the duration that mutant instances exist in the projects lifetime, that has never been studied before. 
%\checkme{In our analysis, for the latent mutants revealed within 365 days, we find that the average lifetime is 72.6 days, and within that time, 103.7 commits were made. These values imply that latent mutants often remain hidden for many changes, imposing significant technical debt on developers. The mutants discarded during the propagation within 365 days, e.g., by deletion, have similar statistics but tend to be more frequently changed. For the mutants never killed, around 76.1\% of them withstand more than a thousand days and commits on average.} %on average 1184.37 days and 1154 commits}
In our analysis, we studied time periods of 365 days and found that the average lifetime of latent mutants is 104 days. This time period involves 136 revisions on average; these values are similar to those overwritten by developer's changes during the inspection and far lower (11 times) than the values of those unkilled to the end.
%\fixme{while average number of revisions that a mutant can exist in the codebase (respective of being killed or not) is  days including 741.8 revisions.}{phrase differntly}}

From a practitioner point of view, the emerging question is how one could identify latent mutants with the information available at a given point in the projects timeline, i.e., without using any future information. We answer this question by using a set of features that are often studied in defect prediction studies and we successfully demonstrate that latent mutants are predictable and they can be predicted with an accuracy of 86\% and a balanced accuracy of 67\% using a simple random forest classifier. 

As such, our latent mutant prediction method falls in the category of the mutant selection methods that have been studied by the mutation testing literature. Specifically, previous work has focused on predicting or identifying killable mutants \cite{titcheu2020selecting} and subsuming mutants \cite{GargODCPT23}. Our key difference from those methods is that we target a subset of mutants that is adding value to the developers rather than the entire set of mutants. Other studies introduced the notions of incremental mutants \cite{ColesLHPV16}, delta-relevant mutants \cite{MaCPH21} and productive mutants \cite{PetrovicI18} that are all linked with a form of incremental development. While important, these techniques target specific code changes and are developer-agnostic. 

Perhaps the closest work to ours, is the fault revealing mutant selection \cite{titcheu2020selecting} that aims at identifying mutants that couple with real faults. While interesting and effective in finding faults, such an approach assumes the availability of a comprehensive set of faults of the environment where mutation testing is to be applied,  to be used as training data. Additionally,  the underlying idea of the fault revealing mutant selection is that those mutants add value to the developers, which is true but does not reflect all cases since developers do not test only for finding faults but also to establish confidence on their code and to prevent introducing faults in the future. This is what latent mutant prediction is doing, it targets cases that are coupled with important future versions.  

This is the first work that studies mutation testing from the software evolution point of view, i.e., the mutants and code co-evolution, identifies the lasting aspect (time lifespan) of the mutants, and comes to an actionable insight about mutants' utility based on the concept of latent mutants. Therefore, we believe that developers can make use of our predictions and target mutants that both last across versions and add them value.  All-in-all our paper makes the following contributions:

\begin{itemize}
  \item We introduce the notion of latent mutants, an important category of mutants that can elicit tests that developers would spent time writing. 

  \item We study the characteristics of latent mutants and inspect the average lifespan of these latent mutants,  demonstrating their added value in testing effort.
  %provide evidence that they couple with latent faults (in X\% of the Defects4J faults).

  \item We provide evidence that latent mutants are predictable, predicted with 86\% accuracy by using code change-related features from the projects history and the used mutant operators. 

\end{itemize}

% section introduction (end)
\section{Background} % (fold)
\label{sec:background}

\subsection{Mutation Testing} % (fold)
\label{sub:mutation_testing}
 
%The main strength of mutation testing lies in its explorability. By simulating various changes a program can undergo during its life cycle, 
Mutation testing guides developers to find unexpected program failures in advance, enabling a thorough testing~\cite{ChekamPTH17}. However, this strength often comes with increased testing efforts. As a result, the identification of useful mutants has always been one of the main interests in mutation testing~\cite{PapadakisK00TH19, GopinathAAJG16}. In most prior studies, this usability is defined based on the mutant's capability to reveal the failure or loop-hole in the current program using the test suite at a given point in time. While these studies successfully reduce the mutation testing cost by selecting fewer valuable mutants, their perspective is often limited to what developers currently have. The value of mutants may change as the program evolves, some becoming useful in later versions (i.e., latent mutants) but for sure the mutants that are discarded (does not exist in any future version) from one version to another are somehow not useful since their utility expires as it only applies at a given revision. 

\subsection{Software Evolution} % (fold)
\label{sub:software_evolution}

%Typical software systems evolve over time,  e.g., due to  software maintenance -- code refactoring, feature implementation and bug fixing. Code changes are either \textit{semantic} or \textit{only syntactic} in nature  \todo{cite}. On the one hand, \textit{Semantic code changes}, such as bug fixing and feature implementation, directly impact the behavior of the program, e.g., its computation and resulting output.  
%On the other hand, syntactic code changes (such as code refactoring,  modularization and program styling). 
Software systems typically evolve over time, e.g., due to software maintenance, undergoing various changes -- code refactoring, feature implementation, bug fixing, etc. 
Code changes can be \textit{semantic} or \textit{only syntactic} in nature. \textit{Semantic code changes}, such as bug fixing and feature implementation, directly impact the behavior of the program, for instance, by changing its computation and resulting output. On the contrary, \textit{syntactic code changes} (such as code refactoring, modularization and program styling), by definition, do not change the program behavior. %However, they may have indirect influence; %, for instance, by affecting the behaviour of developers; prior studies on the relation between fault-introducing changes and refactorings show that these two types of changes are often made together; 
%prior studies show that these syntactic changes are often made together with semantic changes, including those causing failure\todo{~\cite{}}. 

Code changes are particularly common in large code bases with big development teams and a huge number of users. New feature request/implementation and continuous improvement are common in complex and well-maintained software systems. This often results in frequent evolution of the software. Frequent code evolution often makes software testing tasking. This is particularly expensive for mutation testing, where it is computationally expensive to compile and execute multiple mutants across several tests for every program version.  

\subsection{Mutation Testing \& Software Evolution} % (fold)
\label{sub:software_evolution}

Mutation testing of evolving programs is challenging because of high program complexity, scalability of mutation analysis, and the huge cost of change impact analysis~\cite{SunLLLZ15}. More importantly, both semantic and syntactic code changes may impact mutation testing results~\cite{OjdanicSDPT23}. For instance, a statement deletion could be as a result of a bug fixing activity, or program modularization or styling. As illustrated in \autoref{fig:mutant_propagation}, this change can directly influence testing mutants from the previous version on the following version regardless of its purpose: statement deletion at time $t_2$ delete $M_4$ in the program version, resulting in a \textit{killed mutant}, i.e., mutants that no longer exist in the codebase. 
%However, this change directly influence mutation testing: As illustrated in \autoref{fig:mutant_propagation}, statement deletion may influence mutant propagation, thereby resulting in a \textit{dead mutant} (\textit{e.g.,} mutant $M_4$ in the program version at time $t_2$). 

Previously, researchers have studied mutation analysis for evolving systems in a sub-field called\textit{ commit-relevant mutation testing} \cite{MaCPH21}. Albeit, existing studies focus on program changes between two versions of a program, i.e., mutational analysis of the changed code versus unchanged code. In contrast to previous works, this paper investigates the effect of code evolution on mutation testing across the life cycle of a project. The focus of this work is beyond comparing two adjacent versions (or commits) of a software system, investigating how the values of live mutants change as they propagate into future versions of the program.

\section{Overview}

\subsection{Key Insight}

Several researchers have shown that faults may lie dormant in the code base for a long life span before they are discovered, revealed, or fixed~\cite{Cbabral2019icse,AnFonteicse23}. We refer to such faults as \textit{latent} faults. Latent faults may only be revealed due to new test cases or code evolution. For instance, \checkme{Cabral et al. has shown that real faults have a life span of 90 days and up to 11 years} before they are revealed~\cite{Cbabral2019icse}. %Indeed, the time between bug introduction and bug fixing in Defects4J is \todo{XXX} days. 
% 90 days: 1 to 11.5 years (DP). <-> find those in latent fault work 

Drawing from this observation and the fact that mutants are known to couple with real faults, we analogously expect similar latent behaviors in mutants over the lifespan of a software project.  We thus, study \textit{latent mutants} -- mutants that may appear dormant (i.e., live) in the current (or previous) version of the code base,  but reveal faults (i.e., killed) in future code revisions. Particularly,  we examine the prevalence, characteristics, lifespan and predictability of latent mutants.

\begin{figure}[h]
  \centering
   \vspace{-1.0em}
  \includegraphics[width=0.80\textwidth, trim = 10mm 0mm 0mm 0mm]{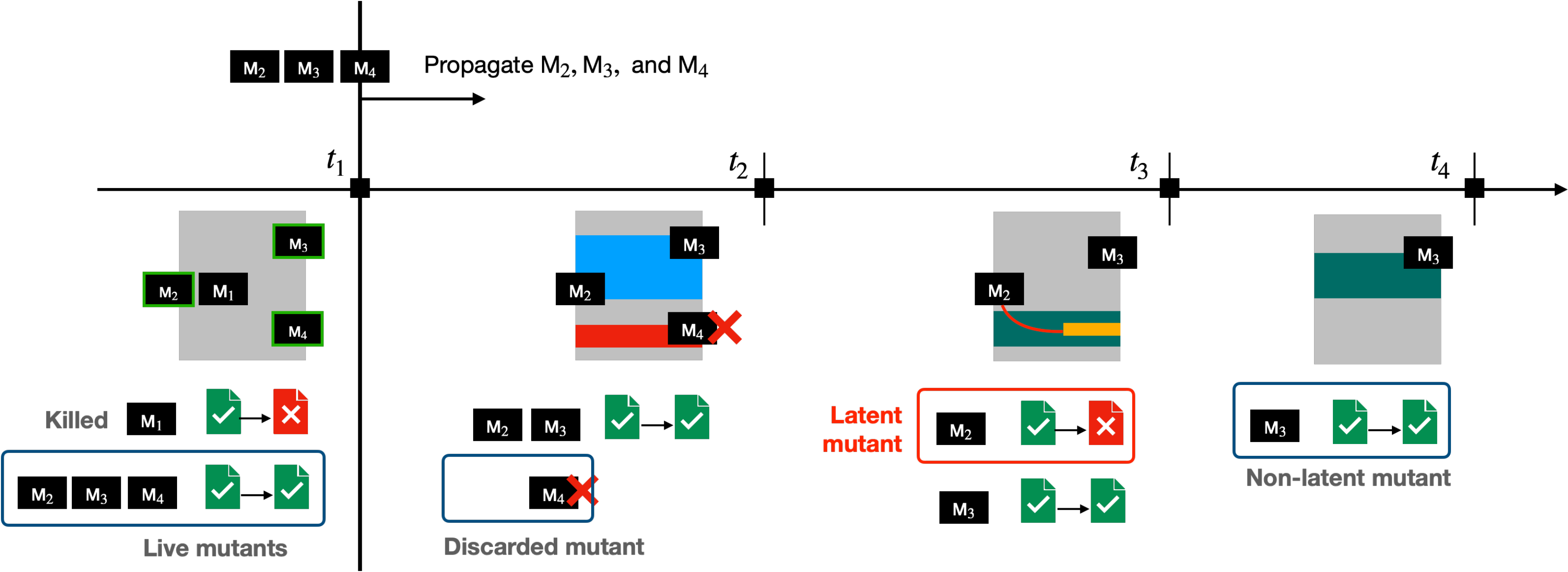}
  \vspace{-1.0em}
  \caption{
  Live mutants at $t_1$, $M_2$, $M_3$ and $M_4$, are propagated to the next versions. 
  Red and blue denote the deletion and the refactoring of code; green refers to semantic changes to the file. $M_2$ is revealed by a new test case at $t_3$, $M_4$ is deleted at $t_2$ and $M_3$ remains undetected. 
  }\label{fig:mutant_propagation}
   \vspace{-1.0em}
\end{figure}

\subsection{Motivating Example}

\cref{fig:mutant_propagation} illustrates the notion of latent mutants and the interplay between mutant lifespan and code evolution. In this example, four mutants ($M_1$, $M_2$, $M_3$, and $M_4$) are injected into the program at $t_1$. The figure shows that three of the injected mutants ($M_2$, $M_3$, and $M_4$,) remain undetected (i.e., not killed) by the available test suite at $t_1$.  We say that these mutants ($M_2$, $M_3$, and $M_4$) are \textit{live} for the program version and test suite at time $t_1$.  Meanwhile,  mutant $M_1$ is killed by the test suite, at time $t_1$. Consequently,  we refer to such mutants (e.g.,  $M_1$) as \textit{killed mutants}. 

Next, let us assume that the live mutants at time $t_1 $ are propagated in future versions of the software at time $t_2$, $t_3$ and $t_4$. Given that the mutated line for mutant $M_4$ is deleted (marked {\color{red}red} in \cref{fig:mutant_propagation}) at time  $t_2$, we \textit{discard} mutant $M_4$. We call such mutants (e.g., $M_4$) \textit{discarded mutants}, as they do not have any impact on future program versions, e.g.,  due to code deletion or overwrite. 

In addition to statement deletion, programs may be refactored (i.e., marked {\color{blue}blue} in \cref{fig:mutant_propagation}) across versions, including the mutated lines of $M_2$ and $M_3$. %These refactorings do not change the program semantics and $M_2$.  
These refactorings do not change the program semantics but may or may not affect the mutants by modifying their locations. Here, $M_2$ and $M_3$ stay undetected by the test suite at time $t_2$, thus are \textit{still live} at $t_2$.  

%In addition to statement deletion, programs may be refactored (i.e., marked {\color{blue}blue} in \cref{fig:mutant_propagation}) across versions, including the mutated lines of $M_2$ and $M_3$. These refactorings do not change the program semantics and $M_2$.  Despite code refactoring, we observe that mutant $M_2$ stays undetected by the test suite at time $t_2$, thus $M_2$  is \textit{still live} at $t_2$.  

However, we observe that mutant $M_2$ is killed by a new test case at time $t_3$. At $t_3$, $M_2$ is not directly changed but killed due to its dependency to the changed lines highlighted in (marked {\color{orange}orange} in \cref{fig:mutant_propagation}). In this work, we call such mutants (e.g., $M_2$) a \textit{latent mutant}. This is because,  this latent mutant was not killed at the point of injection at time $t_1$, but later killed in future program revisions at time $t_3$; note that this definition includes those killed by the changes directly made to them. Latent mutants are important because they may expose faults that were hidden in previous program versions. 

Finally, the mutated statement of mutant $M_3$ is semantically changed (marked {\color{green}green} in \cref{fig:mutant_propagation}) at $t_4$. Still, test cases at $t_4$ fail to detect $M_4$, resulting in $M_4$ being \textit{live} to the end. Such mutants are referred to as \textit{non-latent}. We note that \textit{non-latent mutants} here refer to those that are live throughout the observed lifespan of the software evolution, and they are different from the live mutants that are only live at a specific time (e.g., time of injection).

\section{Terminology and Methodology} % (fold)
\label{sec:terminology_and_methodology}

\subsection{Definition of Terms} % (fold)
\label{sub:definition_of_terms}

%We find three different states of mutants after the propagation in \cref{fig:mutant_propagation}. This section defines different categories of mutants based on these three mutant propagation states and the initial mutation testing results.
We divide mutants into five categories based on the three different propagation states of mutants, shown in \cref{fig:mutant_propagation}, and the initial mutation testing results. 

\noindent
\textbf{Killed Mutants: } A mutant is \textit{killed} if there exists \textit{at least} one test case (in the software's test suite) that \textit{fails} when executed with the mutant. %We say that a test case reveals the fault if it is sufficient to distinguish the observable behavior between the mutant and the original program.  
The goal of mutation testing is to kill all mutants. The number of killed mutants is employed to measure the adequacy of the test suite. Consider \autoref{fig:mutant_propagation} with \textit{four} injected mutants $\{M_1, M_2,  M_3,  M_4\}$. In this example, mutant $M_1$ is \textit{killed} at time $t_1$. %Thus, we say that the \textit{mutation score} at time $t_1$, computed as the ratio of killed mutants over the total number of generated mutants, is \checknumber{0.25}.}{} % do we mention mutation score here? 

\noindent
\textbf{Live Mutants:} This refers to the mutants that remain undetected (i.e., \textit{not killed}) by the existing test suite, i.e., no test case in the current test suite distinguishes the behaviour of the mutants from that of the original program. \autoref{fig:mutant_propagation} shows that three mutants ($M_2,  M_3$ and  $M_4$) \textit{are live} at time $t_1$.\footnote{We call these initial surviving mutants as live to further differentiate with the mutants that survive after the propagation.} 

%\vspace{0.5mm}\noindent\textbf{Propagated Mutants:} In this work, we \textit{propagate} \textit{live mutants} by injecting these mutants into future program versions. %\autoref{fig:mutant_propagation} shows that the two surviving mutants ($M_2,$ and $M_3$) injected in the program version at time $t_1$ were successfully propagated into a future version of the program at time $t_2$. In \autoref{fig:mutant_propagation}, $M_2$, $M_3$, and $M_4$ are propagated mutants, each with a different final state.  
\vspace{0.5mm}
\noindent

%We \textit{propagate} \textit{live mutants}, introducing them to future program versions. In \autoref{fig:mutant_propagation}, $M_2$, $M_3$, and $M_4$ are propagated, each of them belonging to a different group with a different final state.

%\noindent\textbf{Latent Mutants} The mutant that failed to be captured by the current test suite but caught by the future test suite belong to this category. We call these mutants \textit{latent mutants} similar to the naming of latent faults. In \cref{fig:mutant_propagation}, $M_2$ is a \textit{latent mutant} since it is killed by tests at $t_3$ but not by those at $t_1$. 
\noindent\textbf{Latent Mutants:} We investigate if \textit{live mutants} injected at time $t_1$ are \textit{killed} (i.e., reveal faults) in the future, i.e., at time  $t_{1+i}$. We refer to such mutants as \textit{latent mutants}.  Consider \autoref{fig:mutant_propagation}, mutant $M_2$ is a \textit{latent mutant} because it was injected at time $t_1$ and \textit{live} but is \textit{killed} in the future ($t_3$). This paper posits that latent mutants mimic the behavior of real faults, especially faults that are introduced at a certain time, but only revealed in the future, as the program or test suite evolves. 
We believe it is pertinent to study latent mutants because they uncover hidden program failures.  Hence,  this paper focuses on the characteristics and predictability of latent mutants.

%\noindent
%\textbf{Non-latent Mutants:} Non-latent mutants denote the mutants that are killed by none of the future test suites during the propagation. Consider \autoref{fig:mutant_propagation}, mutant $M_3$ is a \textit{non-latent mutant} because it was injected at time $t_1$ and was never killed to the end ($t_4$). We call them non-latent since there is no evidence that supports these mutants to be useful for future testing. 
\noindent
\textbf{Non-Latent Mutants:} Analogously, non-latent mutants refer to mutants that are alive throughtout the lifespan of the project, i.e., they are never killed from the point of injection till the current time or end of the project. %Consider mutant $M_3$ in \autoref{fig:mutant_propagation}, 
In \autoref{fig:mutant_propagation}, mutant $M_3$ is a \textit{non-latent mutant} because it was injected at time $t_1$ and was never killed to the end ($t_4$). We call them non-latent since there is no evidence that supports these mutants to be useful for future testing. 

\noindent
\textbf{Discarded Mutants:} %In this paper, a \textit{propagated mutant} is \textit{dead} if it is deleted or overwritten in future versions. This is due to code evolution, i.e., code refactoring or implementation of new features between both program versions. Thus, dead mutants have no impact on the future version of the program since they are not part of the executable code base. \autoref{fig:mutant_propagation} shows that mutant $M_4$ is a \textit{dead mutant} in program version at time $t_2$ because it is located on a deleted program component (red portion of the program). % due to code refactoring.}
This refers to mutants deleted or overwritten by developers and thus discarded from further inspection. %; code refactoring or implementation of new features between both program versions can be the reason. 
Discarded mutants have no impact on the future version of the program since they no longer exist in the executable code base. In \autoref{fig:mutant_propagation}, mutant $M_4$ is a {discarded mutant} in the program version at $t_2$ because it is located on a deleted program component (red).

\subsection{Code Evolution and Mutant Propagation} % (fold)
\label{sub:code_evolution_and_mutant_propagation}

Software evolves through the changes made by developers; some of them change the program semantics, some only the syntax, and even limited to the style changes. Thus, as depicted in \cref{fig:mutant_propagation}, mutants will likely undergo various changes as they propagate to future versions. To study how the change type affects the final mutant state, mainly their disclosure, we differentiate changes on the mutated lines into three: 1) semantically changed, 2) refactored, and 3) not changed; the last one includes white space and comment changes, as these changes impact neither the syntax nor semantic of the code. The latent mutants can be revealed either by the changes directly made on them or by the changes on dependent code (e.g., the case of $M_2$ in \cref{fig:mutant_propagation}). Combined with the three initial change groups we defined, we categorize latent mutants further into five categories. 

\begin{itemize}
  \item $SC_{C}$. Semantically changed and revealed by changes on the mutated line 
  \item $SC_{NC}$. Semantically changed and revealed by changes on the dependent lines
  \item $RC_{C}$. Refactored and revealed by changes on the mutated line 
  \item $RC_{NC}$. Refactored and revealed by changes on the dependent lines
  \item $NC_{NC}$. Not changed and revealed by changes on the dependent lines
\end{itemize}

%The \textit{mutated line} refers to th e

Different changes can be made on the mutated line. For this categorization of latent mutants, we consider the line \textit{semantically changed} if it was semantically changed even once; we regard the line to be refactored if it only went through code refactorings. Regarding the mutant revelation through the changes on the dependent code, we consider only the changes in the same file. 
%For the dependency between lines, we ..(-> we do not inspect deeply by actually generating the dependency graph (can be future work?)) only limited to the lines within the same file; the mutant check is done if and only if the mutated files is changed (-> threat to validity?) 

% subsection software_evolution (end)

% section background (end)
 
%\input{methodology_terms} 
%\input{study}
%\input{bisection_study}
\section{Experimental Settings}\label{sec:experimental_settings}

%This section presents the experimental settings employed in this work. 

\subsection{Research Questions} % (fold)
\label{sub:research-questions}

%We investigate the following research questions (RQs):% to examine the properties, usefulness and predictability of latent mutants: 

\noindent
\textbf{RQ1: Prevalence: }
What is the prevalence of \textit{live} and \textit{latent mutants}?

\noindent
\textbf{RQ2: Mutation Operations: }
Which mutation operators \textit{frequently} lead to \textit{live} and \textit{latent mutants}? 

\noindent
\textbf{RQ3: Change Features:}
Are there historical change properties (e.g., age, churn and number of developers) that characterize live and latent mutants? 

\noindent
\textbf{RQ4: Lifespan of Latent Mutants :}
How long does it take to kill a latent mutant, i.e., the duration between the injection and its killing time?

\noindent
\textbf{RQ5: Latent Mutant Prediction:} 
Can we predict latent mutants among the live ones?

\subsection{Subjects} % (fold)
\label{sub:subjects}

We use the 13 open-source Java projects of Defects4J v2.0.0~\cite{Just:2014aa} %, a well-known benchmark of real-world faults,
 as our target to study the interactions between mutation testing and code evolution. %\footnote{JfreeChart is excluded due to its version control system (svn), and Mockito and Gson are excluded due to the compilation and testing issues related with the configuration. \fixme{JacksonDatabind is excluded for the time issue.}{}}. 
We use Defects4J because it provides information on bug-fixing commits and the classes modified by the bug-fixes. Additionally, Defects4J offer a list of test classes relevant to the classes modified in bug-fixing commits. Thus, by focusing on these test classes that are likely to execute mutants, we can avoid running tests that are unlikely to cover the mutated code. \cref{tab:subject} shows the general statistics of studied subjects. %\fixme{Out of 647 bug-fixing 

%This paper aims to explore how future software changes can affect the live mutants of mutation testing. Thus, we prefer to mutate the code likely to be actively maintained during the inspection. The bug-fixing commits and the modified classes imply that developers devoted their effort to these classes at this moment (i.e., bug-fixing commits). We thereby use them for when and where to start the investigation. 
%Secondly, Defects4J offers a list of test classes relevant to the classes modified in bug-fixing commits. Thus, by focusing on these test classes that are likely to execute mutants, we can avoid running tests unlikely to cover the mutated code. \cref{tab:subject} shows the general statistics of studied subjects. %\fixme{Out of 647 bug-fixing commits, 547 are studied in the end; 
%We are aware that both mutants and the changes that came after bug-fixing commits, whether in code or tests, can affect the execution of mutated code. Nevertheless, since we have to repeat the test execution for multiple commits during the exploration, we argue that this is an acceptable trade-off. 

\begin{table}[ht] 
% \vspace{-1.0em}
\caption{Subject. All \# are in average. \#$_{modCls}$ and \#$_{relTest}$ refer to the average number of modified classes and relevant test classes. \textit{commit range} is the first and last commit date included in the project repository in Defects4J. \textit{target} denotes those we studied. \#$^{insp}_{commits}$ and \#$^{insp}_{days}$ are the average number of commits and days between bug-fixing and the last commit in the repository}
 \vspace{-1.0em}
\label{tab:subject}
\scalebox{0.7}{\begin{tabular}{lr|rrrr|rrrr}
 \toprule
Proj & \#$_{bug}$ & \#$_{modCls}$ & \#$_{relTest}$ & \#$_{commits}$ & commit range & \#$^{target}_{bug}$ & \#$^{insp}_{commits}$ & \#$^{insp}_{days}$ & \#$_{muts}$ \\
\midrule
Lang & 49 & 1.0 & 7.6 & 3596 & 2002-07-19 -- 2013-10-10  & 49 & 1019.4 & 1170.2 & 16973\\
Math & 104 & 1.1 & 18.5 & 4913 & 2003-05-12 -- 2013-10-16  & 104 & 1972.6 & 987.0 & 28548\\
Time & 20 & 1.2 & 71.0 & 1718 & 2003-12-16 -- 2013-12-04  & 20 & 92.6 & 323.8 & 3086\\
Closure & 139 & 1.2 & 93.5 & 2898 & 2009-11-03 -- 2013-12-13  & 139 & 1180.9 & 598.9 & 33969\\
Cli & 30 & 1.4 & 15.5 & 914 & 2002-06-10 -- 2019-03-25  & 30 & 333.1 & 3196.1 & 2760\\
Compress & 45 & 1.2 & 20.2 & 2682 & 2003-11-23 -- 2019-03-25  & 45 & 1321.1 & 1996.0 & 8371\\
Codec & 18 & 1.6 & 5.8 & 1795 & 2003-04-25 -- 2019-04-23  & 18 & 871.8 & 2551.9 & 3561\\
Collections & 2 & 1.0 & 7.0 & 3091 & 2001-04-14 -- 2019-03-25  & 2 & 265.5 & 1176.3 & 436\\
Csv & 14 & 1.0 & 5.2 & 1290 & 2005-12-17 -- 2019-04-14  & 14 & 440.9 & 1606.8 & 1433\\
JacksonCore & 22 & 1.4 & 46.1 & 1724 & 2011-12-22 -- 2019-04-24  & 22 & 867.0 & 1280.9 & 19536\\
JacksonXml & 3 & 1.0 & 45.7 & 949 & 2010-12-30 -- 2019-05-05  & 3 & 321.3 & 983.7 & 298\\
JxPath & 9 & 1.3 & 18.8 & 598 & 2001-08-23 -- 2018-05-15  & 9 & 146.4 & 3273.8 & 1705\\
Jsoup & 92 & 1.4 & 15.9 & 1261 & 2010-01-17 -- 2019-07-04  & 92 & 466.7 & 1543.5 & 10632\\
    \bottomrule
\end{tabular}}
 \vspace{-1.0em}
\end{table}

\subsection{Mutation Testing} % (fold)
\label{sub:mutation_testing}

We use Pitest~\cite{ColesPitISSTA16}, a popular mutation testing tool for Java, for our mutation analysis.  Pitest provides various mutation operators with different groups. We use the \textit{DEFAULTS} group of mutators, which contains 11 mutators.
resulting in 131,308 mutants generated for 13 projects 
 (\textit{see} \cref{tab:tab1_init_mut_stat}). 
% shows that 131,308 mutants are generated for 13 projects with these operators. 

\subsubsection{Replication}\label{subsub:replication}
Pitest is a tool that operates on byte code. Thus, to propagate the obtained mutants to future versions, we need to convert these bytecode mutants to source code ones. While existing tools, such as Java Decompiler\footnote{https://java-decompiler.github.io/}, can do such a job, the decompiled code often significantly differs from the original source code due to various reasons -- the compiler optimization, bytecode obfuscation, etc. Therefore, instead of using existing decompilation tools, we replicate the mutation based on the information from the mutation testing reports generated by Pitest\footnote{Pitest generates detailed mutation testing reports that include the line number of mutated lines, the used operators, and the description that explain how the mutation has taken place} and its implementation. For this replication, we first identify an AST node to mutate and apply the defined mutation. We select GumTree as a parser for compatibility since it is later used to process changes between files during the propagation; the AST node information of mutants is stored for later propagation. We evaluate whether mutants are successfully replicated by comparing their test pass and fail results with those of the original mutants. Of the 14,747 live mutants  we successfully replicated 12,239 (82.9\%). The first and the second columns of \cref{tab:tab_rq2_mut_status} presents the details. 

\subsubsection{Propagation} % (fold)
\label{subsub:mutant_propagation}

Since our projects employ Git for version control, we also use Git to track the evolution of the projects, mainly the changes made on the mutated files. %With these diff outputs, we propagate the mutants that previously survived, i.e., remain undetected, to the next commit. 
To match the code elements between two program versions, we use GumTree v3.0.0~\cite{GumtreeASE14}. %Various types of changes can be made during the software development. 
\cref{sub:code_evolution_and_mutant_propagation} categorizes code changes into three categories: 1) semantic, 2) refactoring, and 3) style changes, such as those on whitespace and comments, considered as \textit{no-change}. While we use GumTree as the basis to map the AST nodes of code elements between two versions, we further address changes differently depending on their types. %Note that the changes are compared at the file level: i.e., we compare the entire files between versions to check for changes.

%\noindent{\textbf{Style Changes}} 
\par{\noindent\textbf{a) Style Changes:}}\space
To identify whether changes made on a file between two commits are limited to style, we exclude all comment elements from the parsed results of GumTree; we also exclude annotation elements, as they do not affect the program execution. We then reconstruct the file from the parsed outcome, replacing all varying sizes of whitespaces with a single space. If the reconstructed files are the same between two commits, we treat the changes as style changes. 

\par{\noindent\textbf{b) Code Refactorings:}}\space We use RefactoringMiner v2.4.0~\cite{Tsantalis:ICSE:2018:RefactoringMiner} to detect code refactorings in changes. RefactoringMiner uses an AST-based tree-matching algorithm to get the refactoring candidates; it extends ASTDiff APIs of GumTree to generate diffMaps between AST nodes. Eventually, RefactoringMiner generates two types of outputs: AST node diffMaps and code refactoring information of the changed nodes. We use these outputs to propagate and categorise the mutants. 
In this study, we are only interested in specific files where mutants are injected. Thus, for usability, we extend RefactoringMiner to directly support processing a specific file between two arbitrary commits. 

\par{\noindent\textbf{c) Semantic Changes:}}\space We regard the changes to be semantic if they are neither style changes nor code refactorings. Although this definition may not be sound since it includes 
%by including those 
syntactic changes, it is complete. Similar to style changes and code refactorings, we use the diffMaps of GumTree to process the changes. 

With this change information at AST node levels, we propagate the mutants by mapping the information of the mutated nodes between commits. We further compare the values of AST nodes to check whether developers changed the mutated node. In such cases, we \textit{discard} the mutant, as it was overwritten by developers before being revealed. 

During the propagation of a mutant, whenever changes are made to the mutated file, regardless of their type and whether they touch the mutant, we rerun a test suite of the given time. If the mutant introduces a new failing test, we regard it as revealed and thereby \textit{latent}.  

% subsection revealation (end)

\subsubsection{Mutant Status} % (fold)
\label{subsub:mutant_status}

%\cref{sec:terminology_and_methodology} introduces 
We have three states of mutants after the propagation: 1) \textit{latent}, 2) \textit{non-latent}, and 3) \textit{discarded}. In  \cref{fig:mutant_propagation}, non-latent mutants are those that remain live after reaching the end of the inspection. While this might be simply due to the nature of the mutants (e.g. equivalent mutants), it can also be due to the short duration of the inspection. \cref{tab:subject} shows that the maximum inspection time varies between mutants depending on when they are introduced. To avoid introducing noise in the analysis, we use a fixed duration of \textit{N$_{thr}$} days to decide the final mutant status. With this threshold, we redefine the final mutant status after the propagation as follows.  

\begin{itemize}
    \item \textit{non-latent}: mutants that remain uncaught at least more than \textit{N$_{thr}$} days
    \item \textit{discarded}: mutants that are overwritten/deleted by developers' changes during the inspection 
    \item \textit{latent}: mutants that are revealed by new failing tests within \textit{N$_{thr}$} days
\end{itemize}%

We apply this threshold to obtain latent mutants and to exclude those mutants that take too long to be revealed. For \textit{discarded} mutants, we use all of them to differentiate this type of mutants from non-latent mutants further. In this study, we set \textit{N$_{thr}$} to 365 days.

\subsection{Change Feature Collection} % (fold)
\label{sub:change_feature_collection}

We inspect the historical properties of the mutated code elements, mainly the metadata that describes the past change trends of code elements. We select three types of historical features, i.e., \textit{age}, \textit{churn} and \textit{developers}, shown to be effective in describing the past change tendencies of code elements concerning their likelihood of causing test failures~\cite{McI:2018:tse, Kamei2013aa,Sohn2021saner}. 

\begin{itemize}
    \item \textit{Age} measures how long the code element has been in the codebase. We measure two age metrics, i.e., the minimum ($age_{min}$) and the maximum ($age_{max}$): the former computes the time interval from the last change, whereas the latter calculates the interval from the first introduction of the code element. 
    \item \textit{Churn} evaluates how frequently changed the code element is, counting the number of past changes made to the code element. 
    \item \textit{Developers} is the number of unique developers that modified the code element. 
\end{itemize}

%x\cref{sub:rq1} introduces 
We compute these four features per line ($l$) and method ($m$), as the line granularity can be too fine-grained to capture the general trend. As a result, eight features are investigated in total: $age^l_{min}$, $age^l_{max}$, $churn^l$, $n^l_{dev}$ with the line granularity and $age^m_{min}$, $age^m_{max}$, $churn^m$, $n^m_{dev}$ with the method granularity. We compare the trend of these feature metrics between live and killed mutants. 

Overall, we employ eight historical features to analyse the characteristics of the mutants. We collect these features using Git and javalang v.0.13.0\footnote{https://github.com/c2nes/javalang}, a lexer and parser for Java 8. 
% subsection change_feature_collection (end)

\subsection{Model Training} % (fold)
\label{sub:model_training}

To answer \textbf{RQ5}, we train a simple classifier that predicts the final labels of live mutants, i.e., latent, non-latent, and discarded.  Model features are the eight historical change features and the mutation operators chosen to study the mutant characteristics in previous research questions. For the training algorithm, we select Random Forest (RF), which has often been employed to predict the defect-proneness of the code, i.e., the likelihood of causing a program failure, using the historical change features~\cite{McI:2018:tse,Kamei2016xr,Sohn2021saner}. As the primary objective of this study lies in exploring the characteristics of latent mutants, we further investigate the feature importance of the trained model.

We use five-fold cross-validation to split training and test data. For Random Forest classifiers, we employ scikit-learn~\cite{sklearn_api} and the default parameters provided for the model training. We repeat the experiments ten times to reduce the randomness involved in RF and report the average results. 

\subsection{Model Performance} % (fold)
\label{sub:model_performance}
We evaluate the performance of our model using accuracy and balanced accuracy, per class. We also compute the Mean Average Precision (MAP). MAP is calculated as the mean of the average precision of all revisions, i.e., fixed commits, for each project. Here, we regard the latent mutants as the target to predict and compute the precision at each point in the list of mutants ranked in descending order of their likelihood to be latent. In the end, we define MAP as follows. 
%}

\vspace{-0.7em}
\begin{equation}%
\scalebox{0.95}{  
MAP = $\frac{1}{\#_{rev}}\sum_{i = 1}^{M} \frac{Precision(i) \times isLatent(i)}{\# \text{of latent mutants}}\text{,  } Precision(i) = \frac{\#\text{ of latent mutants within the top i}}{i}$
}
\end{equation}
\vspace{-0.7em}

%\revise{
\noindent \#$_{rev}$ refers to the number of revisions, $M$ to the number of mutants per revision. Thus, if a revision contains only latent mutants or none, we skip the computation and mark the case with "--". MAP simulates the use case where developers have a set of mutants and want to identify the latent ones.

% subsection evaluation_metrics (end)
\subsection{Implementation and Environment} % (fold)
%\subsection{Implementation Details and Platform} % (fold)
\label{subsub:environment}

All experiments were conducted on a single threaded process on a compute server with 28 cores and 128 GB of RAM running an Skylake CPU.  %2.6GHz with \todo{xXX}  virtual cores,
Mutant generation and propagation are implemented in Python 3.8.6. %The replication package and the collected data are publicly available in \url{https://anonymous.4open.science/r/latent_mutExploration-CE27}

%\todo{LOC of implementation of methodology}
%\todo{LOC of data analysis}
%\todo{compute environment used,  server details }
%\todo{tools and library used in the implementation}
%\todo{anonymous link to source code and experimental/replication data}

% subsubsection environment (end)

\section{Results}\label{sec:result_analysis}

\subsection{RQ1: Prevalence}
\label{sub:rq1_prevalence}
%We inspect the prevalence of live mutants using thirteen subject programs, 547 bug-fixing commits %and up to 1972.6 program versions per bug, 
%from Defects4J (\textit{see} \cref{tab:subject}). 
\cref{tab:tab1_init_mut_stat} and \cref{tab:tab_rq2_mut_status} show the prevalence of live mutants and latent mutants, respectively. 

\begin{table}[bt!] 
\vspace{-0.6em}
\caption{Prevalence of Live and Killed mutants. ``Others'' include No coverage, Time out, Non viable, Memory error, and Run error. \textit{avg} and \textit{agg} refer to the average and the aggregation across revisions. %``Total'' is computed with all mutants. 
}\label{tab:tab1_init_mut_stat}
\vspace{-1.0em}
\scalebox{0.65}{
\begin{tabular}{lr|rr|rr|rr|rr} %|rr}
    \toprule
     & & \multicolumn{2}{c|}{All (\#)} & \multicolumn{2}{c|}{Live (\%)}  & \multicolumn{2}{c|}{Killed (\%)} & \multicolumn{2}{c}{Others (\%)} \\
Proj & \#$_{rev}$ & avg & agg & avg & agg & avg & agg & avg & agg \\
\midrule
Lang & 49 & 346.39 & 16973 & 8.18 & 7.12 & 56.4 & 56.04 & 35.42 & 36.84 \\
Math & 104 & 274.5 & 28548 & 11.74 & 10.05 & 65.93 & 61.31 & 22.33 & 28.63 \\
Time & 20 & 154.3 & 3086 & 9.62 & 8.85 & 83.7 & 84.54 & 6.67 & 6.61 \\
Closure & 139 & 244.38 & 33969 & 8.95 & 10.07 & 72.14 & 74.26 & 18.91 & 15.66 \\
Cli & 30 & 92.0 & 2760 & 8.94 & 9.93 & 86.16 & 85.25 & 4.91 & 4.82 \\
Compress & 45 & 186.02 & 8371 & 15.08 & 16.9 & 71.98 & 67.69 & 12.94 & 15.41 \\{}
Codec & 18 & 197.83 & 3561 & 11.64 & 10.11 & 80.74 & 83.46 & 7.62 & 6.43 \\
Collections & 2 & 218.0 & 436 & 11.63 & 14.91 & 28.1 & 19.72 & 60.26 & 65.37 \\
Csv & 14 & 102.36 & 1433 & 8.63 & 12.35 & 84.65 & 80.88 & 6.72 & 6.77 \\
JacksonCore & 22 & 888.0 & 19536 & 16.72 & 14.78 & 58.47 & 61.61 & 24.8 & 23.61 \\
JacksonXml & 3 & 99.33 & 298 & 7.89 & 8.39 & 56.16 & 55.37 & 35.95 & 36.24 \\
JxPath & 9 & 189.44 & 1705 & 10.33 & 9.21 & 72.64 & 72.43 & 17.02 & 18.36 \\
Jsoup & 92 & 115.57 & 10632 & 14.74 & 15.17 & 74.98 & 73.78 & 10.28 & 11.05 \\
\midrule
Total & 547 & 240.05 & 131308 & 11.33 & 11.23 & 71.02 & 67.3 & 17.64 & 21.47 \\
    \bottomrule
\end{tabular}}
\vspace{-1.0em}
\end{table} 

\vspace{0.5mm}
\noindent
\textbf{Prevalence of Live Mutants:}
%the initial results of mutation testing. 
Our results show that \textit{live mutants are prevalent: All inspected projects contain live mutants with about 11\% on average per project
}. 
There are up to $\approx$17\% of live mutants on avaerage per project (\textit{see} \textsc{JacksonCore} \cref{tab:tab1_init_mut_stat}). 
%in some projects (). 
%with  } 
\cref{tab:tab1_init_mut_stat} also shows that one in nine (11\%) mutants are live. Specifically, there were 14,747 live mutants out of the 131,308.  Subsequently,  
%In total, 131,308 mutants are generated. Among them, 11.23\% of them (14747) remain undetected while 
71.02\% of all mutants are killed,  i.e., they were detected by the test suite.  
21.47\% of the mutants were 
%includes mutants 
killed by other factors, (e.g., time-out) %,  
or were 
% with the mutants 
not executed by the test suite (\textit{see} \cref{tab:tab1_init_mut_stat}).\footnote{Since we are interested in how the mutants executed but failed to be detected differs from those killed,  we consider the mutants in \textit{Others} out-of-scope.  Thus, the rest of our study focuses only on the mutants explicitly labeled as \textit{Live} and \textit{Killed}. }

%Overall,  the 
%prevalence of live mutants 
%suggests \revise{
%that either the available test suites are not sufficient to expose them or they are not worth targeted.  }
%Indeed,  our result shows that  live mutants represent a reasonable portion
%(one out of every nine) of injected mutants. 
%Hence,  it is important to study the characteristics of latent mutants
%since they 
%, especially if they are latent, i.e., 
%potentially hide faults. 
%live mutants are prevalent
%\todo{.what is the implication of this result}.  

\begin{result}
All studied projects contain live mutants,  %with 
 with up to $\approx$17\% of mutants being live per project: \\
 Live mutants are prevalent, %All projects contain live mutants 
 about one in nine injected mutants are live mutants. 

\end{result}

\begin{table}[ht] 
\vspace{-0.8em}
\caption{
%Details of the 
Prevalence of Latent, Non-latent and Discarded Mutants.  
%Mutant Propagation Status. 
\#$^{prop}_{rev}$ and \#$^{prop}_{mut}$ show how the number of revisions and propagated mutants change through preprocessing: from right to left, the numbers refer to those with at least one propagated live mutants, those replicated, and those remaining after the threshold of 365 days for the propagation status; %76.1\% and 57.3\% of non-latent and latent mutants remain after the filtering with the threshold. 
 %Out of 12,239 propagated mutants from, 10383 mutants (84.8\%) remain after the exclusion of mutants with the threshold of 365 days; 
With this threshold of 365 days, 76.1\% and 57.3\% of non-latent and latent mutants remain. 
\textit{average} is the average per revision, and \textit{agg} contains the aggregated value per project. 
The last row, \textit{Total}, contains the percentage computed with all generated mutants, and \textit{avg} per revision. 
}\label{tab:tab_rq2_mut_status}
\vspace{-1.0em}
\scalebox{0.65}{
\begin{tabular}{lrr|r|rr|rr|rr}
\toprule
%proj & & \#$_{rev}$ & \#$_{mut}$ (avg.) & average & total & average & total & average & total \\
& \multicolumn{2}{c|}{\#$^{prop}_{rev}$} & \#$^{prop}_{mut}$ & \multicolumn{2}{c|}{Latent (\# (\%))} & \multicolumn{2}{c|}{Non-latent (\# (\%))} & \multicolumn{2}{c}{Discard (\# (\%))} \\
%proj & init & thr$_{365}$/valid/all &  & \#$_{mut}$ & average & total & average & total & average & total \\
%proj & \# & \#$^{v,365}_{w.l}$/\#$^{v}_{w.l}$/\#$^{w.l}$ & \#$_{mut}$ (avg.) & average & total & average & total & average & total \\
proj & \#$_{rev}$ & & \#$_{mut}$ & average & agg & average & agg & average & agg \\
\midrule 
Lang & 49 & \textbf{27}/30/33& \textbf{833}/1025/1209& 1.4 (9.4\%) & 39 (4.7\%) & 22.9 (72.3\%) & 618 (74.2\%) & 6.5 (18.3\%) & 176 (21.1\%)\\
Math & 104 & \textbf{79}/89/92& \textbf{1442}/2099/2870& 1.7 (6.0\%) & 136 (9.4\%) & 10.5 (63.1\%) & 831 (57.6\%) & 6.0 (30.9\%) & 475 (32.9\%)\\
Time & 20 & \textbf{10}/19/19& \textbf{79}/185/273& 0.5 (12.8\%) & 5 (6.3\%) & 7.3 (86.0\%) & 73 (92.4\%) & 0.1 (1.2\%) & 1 (1.3\%)\\
Closure & 139 & \textbf{100}/118/118& \textbf{2242}/2866/3421& 0.3 (4.0\%) & 33 (1.5\%) & 16.7 (68.0\%) & 1668 (74.4\%) & 5.4 (28.0\%) & 541 (24.1\%)\\
Cli & 30 & \textbf{22}/25/25& \textbf{232}/237/274& 0.5 (3.7\%) & 10 (4.3\%) & 8.4 (89.4\%) & 184 (79.3\%) & 1.7 (6.9\%) & 38 (16.4\%)\\
Compress & 45 & \textbf{43}/44/44& \textbf{1089}/1201/1415& 0.6 (3.4\%) & 26 (2.4\%) & 12.7 (58.7\%) & 544 (50.0\%) & 12.1 (37.8\%) & 519 (47.7\%)\\
Codec & 18 & \textbf{16}/16/18& \textbf{271}/282/360& 0.8 (2.6\%) & 13 (4.8\%) & 9.2 (54.8\%) & 148 (54.6\%) & 6.9 (42.6\%) & 110 (40.6\%)\\
Collections & 2 & \textbf{2}/2/2& \textbf{4}/4/65& 0.0 (0.0\%) & 0 (0.0\%) & 2.0 (100.0\%) & 4 (100.0\%) & 0.0 (0.0\%) & 0 (0.0\%)\\
Csv & 14 & \textbf{11}/11/12& \textbf{143}/152/177& 0.1 (9.1\%) & 1 (0.7\%) & 10.5 (63.1\%) & 115 (80.4\%) & 2.5 (27.8\%) & 27 (18.9\%)\\
JacksonCore & 22 & \textbf{21}/22/22& \textbf{2635}/2645/2888& 2.4 (1.2\%) & 51 (1.9\%) & 15.7 (37.6\%) & 330 (12.5\%) & 107.3 (61.2\%) & 2254 (85.5\%)\\
JacksonXml & 3 & \textbf{3}/3/3& \textbf{17}/17/25& 0.0 (0.0\%) & 0 (0.0\%) & 1.3 (26.7\%) & 4 (23.5\%) & 4.3 (73.3\%) & 13 (76.5\%)\\
JxPath & 9 & \textbf{9}/9/9& \textbf{135}/135/157& 0.0 (0.0\%) & 0 (0.0\%) & 13.6 (95.8\%) & 122 (90.4\%) & 1.4 (4.2\%) & 13 (9.6\%)\\
Jsoup & 92 & \textbf{79}/85/85& \textbf{1261}/1391/1613& 0.6 (8.7\%) & 50 (4.0\%) & 5.3 (41.6\%) & 415 (32.9\%) & 10.1 (49.7\%) & 796 (63.1\%)\\
\midrule
Total & 547 & \textbf{422}/473/482& \textbf{10383}/12239/14747& 0.9 (5.5\%) & 364 (3.5\%) & 12.0 (61.3\%) & 5056 (48.7\%) & 11.8 (33.1\%) & 4963 (47.8\%)\\
\bottomrule
\end{tabular}}
\vspace{-0.5em}
\end{table}

\vspace{0.5mm}
\noindent
\textbf{Prevalence of Latent Mutants:} To determine the prevalence of latent mutants, we employed 10,383 live mutants from our previous analysis of killed and live mutants. We excluded 1,856 ambiguous live mutants in our analysis, e.g., because they are beyond our study period of 365 days. Note that we have propagated all 12,239 mutants but focus on this subset to avoid premature conclusions, taking those certain as discussed in \cref{subsub:mutant_status}. 

Results show that \textit{most projects (77\%, 10 out of 13 inspected projects) have at least one latent mutant per version.  There are up to 136 latent mutants per project. } The last row (i.e., \textit{Total}) of \cref{tab:tab_rq2_mut_status} further shows that \textit{about one in 25 live mutants are latent mutants after propagation on average.} 
In our study, latent mutants represent about 3.5\% of live mutants and \checknumber{0.28\%} of all injected mutants.  
Specifically,  we found that 3.5\% (364 out of 10,383) of live mutants are latent, i.e.,  killed within 365 days after their injection.  
Within 365 days, about 48.7\% of live mutants are non-latent (i.e., still live) and another 47.8\% 
%survive
% or 
are discarded after the propagation because they appear on modified or deleted program statement.  
%Overall,  t
This result shows that mutants behave similarly to real faults: Like real faults, mutants can be latent in evolving software systems, suggesting the existence of hidden faults in them. 
%This suggests that most projects contain hidden faults that are only discovered in future revisions. 
Thus, it is crucial to study the characteristics and devise effective means to identify and predict latent mutants. 
%Considering that these mutants are likely linked to latent faults, this 
%\todo{implication of results}
% The exact percentage of the final status varies depending on the project. However, overall, only a small percentage of mutants are identified as latent across all projects.

\begin{result}
Latent mutants are highly prevalent among our examined software projects: 
77\% (10 out of 13) of inspected projects contain latent mutants and 
%between one to  
%latent mutants with 
there are up to 
136 latent mutants 
per project.    

\end{result}

\begin{table}[ht] 
\vspace{-0.5em}
\caption{
Details of Mutation operators (aka mutators) 
%propeties 
leading to Live Mutants at mutant injection time.  
Each value is reported in the percentage of mutant liveness when the mutator in the column is used: $\frac{n_{surv}}{n}$, $n$ is the number of mutants generated by the mutator. Highlighted in blue refers to the most effective, and green refers to those within top three.
%Those with the highest ratios are in bold, and underlined means the operators to be within the top five operators with the highest ratio of live mutants when used. 
}\label{tab:tab_rq1_surv_mutop}
\vspace{-1.0em}
\scalebox{0.7}{
\begin{tabular}{l|rrrrrrrrrrr}
    \toprule

\textit{Perc (\%)} & MATH & CB & INCR & IN & NC & VMC & PRET & ERET & BFRET & BTRET & NRET \\
\midrule

Lang & \cellcolor{green!25}13.1 & \cellcolor{blue!25}\textbf{32.4} & 1.2 & 3.1 & 2.2 & \cellcolor{green!25}6.3 & 5.0 & 1.5 & 0.0 & 1.9 & 0.8\\
Math & 8.9 & \cellcolor{blue!25}\textbf{31.6} & 1.1 & \cellcolor{green!25}10.3 & 6.0 & \cellcolor{green!25}16.5 & 3.5 & 0.9 & 6.1 & 9.6 & 0.3\\
Time & 10.9 & \cellcolor{blue!25}\textbf{30.8} & 8.1 & \cellcolor{green!25}19.4 & 6.4 & \cellcolor{green!25}13.3 & 1.0 & 4.6 & 2.6 & 7.7 & 5.2\\
Closure & 10.1 & \cellcolor{blue!25}\textbf{26.4} & 8.0 & 0.0 & 3.6 & \cellcolor{green!25}19.4 & 5.0 & 3.6 & 6.0 & 7.2 & \cellcolor{green!25}18.7\\
Cli & 14.3 & \cellcolor{blue!25}\textbf{41.6} & 4.3 & - & 2.6 & 8.8 & 4.3 & 3.7 & 5.3 & \cellcolor{green!25}16.7 & \cellcolor{blue!25}\textbf{41.6}\\
Compress & \cellcolor{green!25}17.1 & \cellcolor{blue!25}\textbf{42.2} & 6.0 & 0.0 & 9.4 & \cellcolor{green!25}24.6 & 12.5 & 15.3 & 8.2 & 7.9 & 0.8\\
Codec & \cellcolor{green!25}11.9 & \cellcolor{blue!25}\textbf{40.3} & 2.4 & - & 7.1 & 3.8 & 0.0 & \cellcolor{green!25}8.5 & 3.1 & 2.7 & 3.5\\
Collections & 14.3 & \cellcolor{green!25}30.4 & 0.0 & - & \cellcolor{green!25}18.3 & 10.7 & \cellcolor{blue!25}\textbf{50.0} & 12.5 & 0.0 & 9.1 & 8.6\\
Csv & \cellcolor{blue!25}\textbf{69.7} & \cellcolor{green!25}31.4 & 0.0 & - & 5.2 & \cellcolor{green!25}14.8 & 12.5 & 1.6 & 0.0 & 0.0 & 0.0\\
JacksonCore & \cellcolor{green!25}9.5 & \cellcolor{blue!25}\textbf{50.9} & 9.2 & 0.0 & \cellcolor{green!25}12.1 & 9.1 & 4.4 & 2.2 & 8.1 & 2.3 & 5.6\\
JacksonXml & 0.0 & \cellcolor{blue!25}\textbf{33.3} & 0.0 & - & 4.8 & 10.0 & 0.0 & 3.4 & 0.0 & \cellcolor{green!25}14.3 & \cellcolor{green!25}25.0\\
JxPath & \cellcolor{green!25}10.3 & \cellcolor{green!25}41.2 & 0.0 & - & 6.4 & \cellcolor{blue!25}\textbf{42.5} & 7.5 & 9.1 & 10.0 & 7.4 & 1.6\\
Jsoup & 10.7 & \cellcolor{blue!25}\textbf{38.4} & 1.5 & - & 5.3 & \cellcolor{green!25}30.4 & 9.7 & 4.1 & 10.0 & 9.9 & \cellcolor{green!25}23.0\\
\midrule
Total & 10.5 & \cellcolor{blue!25}\textbf{37.4} & 4.2 & 9.9 & 5.7 & \cellcolor{green!25}18.2 & 4.9 & 3.2 & 5.8 & 6.6 & \cellcolor{green!25}11.6\\
\bottomrule
TopFreq & \cellcolor{green!25}6 & \cellcolor{blue!25}\textbf{14} & 0 & 2 & 2 & \cellcolor{green!25}9 & 1 & 1 & 0 & 2 & 5\\
    \bottomrule
\end{tabular}}
\vspace{-0.5em}
\end{table} 

\subsection{RQ2: Mutation Operations}
\label{sub:rq2_mutation_operations}
Let us examine the characteristics of latent mutants versus live/killed mutants w.r.t. the mutation operators used in generating the mutants. \cref{tab:tab_rq1_surv_mutop,tab:tab_rq1_kill_mutop} illustrates 
the proportion of live mutants and killed mutants
when using each mutation operator, respectively.  
In addition,  \cref{tab:tab_rq2_rv_mut_mutop} illustrates the mutators for latent mutants. 
%\todo{table for latent mutants}
For all three tables,  the mutation operator that leads to the highest ratio of live, killed or latent mutants are highlighted in {\color{blue}blue}, the remaining two operators within the top three are highlighted in {\color{green}green}. The last row, (``\textit{TotalFreq}'') presents the frequency of mutators ranked within the top three across projects\footnote{The values with the same rank are all highlighted (e.g., the last row of \cref{tab:tab_rq1_kill_mutop})}.

\begin{table}[ht] 
\vspace{-0.5em}
\caption{
%Mutators and Killed Mutant. 
Details of Mutation operators (aka mutators) leading to Killed Mutants at mutant injection time. 
Each value is reported in the percentage of mutant kill when the mutator in the column is used: $\frac{n_{kill}}{n}$, $n$ is the number of mutants generated by the mutator 
}\label{tab:tab_rq1_kill_mutop}
\vspace{-1.0em}
\scalebox{0.7}{
\begin{tabular}{l|rrrrrrrrrrr}
    \toprule
\textit{Perc (\%)} & MATH & CB & INCR & IN & NC & VMC & PRET & ERET & BFRET & BTRET & NRET \\
\midrule
Lang & 51.2 & 33.8 & 56.2 & 37.5 & \cellcolor{green!25}62.4 & 44.4 & \cellcolor{blue!25}\textbf{68.1} & \cellcolor{green!25}61.6 & 56.2 & 55.0 & 54.1\\
Math & 63.0 & 39.5 & \cellcolor{green!25}68.1 & \cellcolor{blue!25}\textbf{72.1} & 66.0 & 51.4 & 64.9 & 62.7 & 51.5 & 47.0 & \cellcolor{green!25}71.8\\
Time & 80.8 & 62.8 & 78.4 & 80.6 & \cellcolor{green!25}89.8 & 70.3 & \cellcolor{blue!25}\textbf{93.8} & 87.2 & 89.5 & 72.3 & \cellcolor{green!25}89.8\\
Closure & 78.3 & 58.3 & 71.7 & 31.6 & \cellcolor{green!25}83.0 & 62.5 & 51.7 & \cellcolor{green!25}82.9 & \cellcolor{blue!25}\textbf{83.1} & 79.5 & 66.6\\
Cli & 71.4 & 52.2 & 87.0 & - & \cellcolor{blue!25}\textbf{94.8} & 85.9 & 73.9 & \cellcolor{green!25}92.0 & \cellcolor{green!25}89.5 & 80.2 & 52.6\\
Compress & 68.0 & 47.8 & \cellcolor{green!25}89.7 & \cellcolor{blue!25}\textbf{100.0} & 78.5 & 56.6 & 54.9 & 61.0 & 42.4 & 53.2 & \cellcolor{green!25}86.2\\
Codec & 85.5 & 54.4 & 86.5 & - & 85.2 & \cellcolor{green!25}90.9 & \cellcolor{blue!25}\textbf{95.1} & 83.0 & 81.5 & 80.2 & \cellcolor{green!25}86.9\\
Collections & 7.1 & 8.7 & \cellcolor{blue!25}\textbf{66.7} & - & 18.3 & 7.1 & \cellcolor{green!25}50.0 & \cellcolor{green!25}62.5 & 33.3 & 18.2 & 23.4\\
Csv & 25.8 & 52.3 & 76.9 & - & 89.6 & 77.0 & 78.1 & \cellcolor{green!25}95.3 & \cellcolor{blue!25}\textbf{100.0} & 86.3 & \cellcolor{green!25}98.0\\
JacksonCore & 69.2 & 35.5 & \cellcolor{green!25}73.0 & \cellcolor{blue!25}\textbf{100.0} & \cellcolor{green!25}71.1 & 41.5 & 68.4 & 69.5 & 51.9 & 52.5 & 69.5\\
JacksonXml & 11.1 & 44.4 & \cellcolor{blue!25}\textbf{100.0} & - & \cellcolor{green!25}73.0 & 50.0 & 28.0 & 37.9 & \cellcolor{green!25}60.0 & 28.6 & 50.0\\
JxPath & 66.7 & 33.8 & 54.5 & - & \cellcolor{green!25}78.1 & 42.5 & 46.2 & 68.9 & \cellcolor{green!25}81.4 & 73.6 & \cellcolor{blue!25}\textbf{86.0}\\
Jsoup & 77.5 & 56.7 & \cellcolor{green!25}88.1 & - & \cellcolor{blue!25}\textbf{89.0} & 53.1 & 77.7 & \cellcolor{green!25}83.7 & 80.2 & 70.2 & 59.2\\
\midrule
Total & 65.8 & 41.6 & 71.8 & 70.7 & \cellcolor{blue!25}\textbf{76.2} & 57.5 & 64.3 & \cellcolor{green!25}73.8 & \cellcolor{green!25}75.1 & 70.3 & 67.5\\
\bottomrule
TopFreq & 0 & 0 & \cellcolor{green!25}6 & 3 & \cellcolor{blue!25}\textbf{9} & 1 & 4 & \cellcolor{green!25}7 & \cellcolor{green!25}6 & 0 & \cellcolor{green!25}6\\
\bottomrule
\end{tabular}}
\vspace{-0.5em}
\end{table}  

\vspace{0.5mm}
\noindent
\textbf{Live Mutants vs. Killed Mutants:} 
%
%\cref{tab:tab_rq1_surv_mutop,tab:tab_rq1_kill_mutop} show the proportion of surviving and killed mutants when using each operator, respectively. The operator that leads to the highest ratio of surviving/killed mutant are highlighted in blue; those within the top five are highlighted in green; the last row, \textit{TotalFreq}, presents the frequency of mutators within the top five across projects. 
%Overall, 
Our evaluation results show that \textit{mutators that impact the program's execution paths (e.g.,  NC and ERET) are more likely to lead to killed mutants, while mutators that do not change the program execution path (e.g.,  CB and VMC) lead to live mutants. }
Comparing killed mutants and live mutants, \textit{we found that some mutators (aka mutation operators) are more prone 
%that the mutators that lead 
to killed mutants and 
%are different form thos that leads to 
live mutants}. 
\cref{tab:tab_rq1_surv_mutop,tab:tab_rq1_kill_mutop} show the proportion of live mutants and killed mutants for each operator, respectively.   
Notably,  NC and CB are the mutators most likely to lead to a killed mutant and live mutant respectively. 
On the one hand,  the mutators  that are less likely to change the initial program execution path tend to generate more live mutants, i.e., mutants that are undetected by the available test suite,  as shown in \cref{tab:tab_rq1_surv_mutop}.  On the other hand,  we observed that 
the mutators with a higher mutant-kill ratio (i.e., leading to killed mutants) are those that directly affect the execution path. 

%Concretely,
For instance,  both Conditionals Boundary Mutator (CB) and Negate Conditionals (NC) operators mutate the conditionals.  However,  CB that makes subtle changes on the condition boundary (e.g., $<$ to $\leq$) has the highest likelihood of live mutants whereas NC that explicitly changes the execution path by negating the condition results in the highest mutant kills.  Similarly, the Void Method Call (VMC) operator, which mutates method calls to void methods, is more likely to generate live mutants than other mutators (such as Empty Returns (ERET)) that mutate the method calls to non-void methods.  
%Overall,  t
This result 
%comparisons 
demonstrates that mutation operators (mutators) play  significant roles in the likelihood of a mutant getting killed or living. 

%the importance of 
%the role of mutators on the mutant 
% survivals and kills,  \fixme{conforming with the findings on .. }{cite} % JJ: not sure whether there are some studies that explicitly 

\begin{result}
Mutation operators 
%(e.g., NC and ERET) 
that impact a program's execution path are more likely to lead to killed mutants. In contrast, operators 
% that those 
that do not impact
% with 
%the 
a program's 
execution path often lead to live mutants. 
%Consequently,  killed mutants 
\end{result}

\vspace{0.5mm}
\noindent
\textbf{Latent Mutants vs.  Killed/Live Mutants:}
We found that \textit{the mutation operators that are most likely to lead to 
latent mutants are drawn from the top operators for killed mutants and live mutants} 
The most potent mutators for latent mutants are drawn from the most effective mutators from both live mutants and killed mutants.  
As an example,  
negate conditionals (NC) is the most effective mutator for generating both 
killed mutants and 
latent mutants (\textit{cf.} \cref{tab:tab_rq1_kill_mutop} and \cref{tab:tab_rq2_rv_mut_mutop}).  
In contrast,  the MATH mutator which is the third most effective mutator for live mutants is also the second most effective mutator for latent mutants
% and also the third most effective for live mutants 
(\textit{cf. } \cref{tab:tab_rq1_surv_mutop} and \cref{tab:tab_rq2_rv_mut_mutop}).  
This result implies that latent mutants share mutator properties with both 
live mutants and killed mutants.  
This demonstrates the uniqueness of latent mutants (vs.  killed/live mutants) and the potential difficulty of predicting latent mutants. 
%suggests that predicting latent mutants is non-trivial.  
In particular, one can not reliably identify latent mutators using either the mutators of killed mutant or live mutants. 

%... 

%\revise{
%\fixme{}{include only five ops?} % Fig 3 might be too large and contain unimportant information as our focus is not on surviving and dead mutants. So, drop and include only the top five mostly used operators? 
%The majority of propagated mutants either survive to the end or become lost, i.e., dead, during the propagation, as shown in \cref{tab:tab_rq2_mut_status}. \cref{fig:rq2_mut_op_pie} compares the proportions of latent, \fixme{surviving}{}, and dead mutants for the eight mutators selected from the set of effective mutators for initial surviving (3) and killed mutants (5). The pie charts in the figure show that the usage of different mutators leads to different characteristics of mutants; for instance, VMC tends to generate more mutants that remain undetected to the end. 
%}

\begin{result}
Mutators leading to live or killed mutants are not a reliable proxy for identifying
%/predicting 
latent mutants:
Mutators that lead to latent mutators are drawn from both 
the most prominent mutators of both live mutants and killed mutants. 
%This suggests that predicting latent mutators is no 
\end{result}

\begin{figure}[h]
\centering
\vspace{-0.5em}
\begin{subfigure}{0.18\textwidth}
    \includegraphics[width=\textwidth, trim = 0mm 10mm 0mm 10mm]{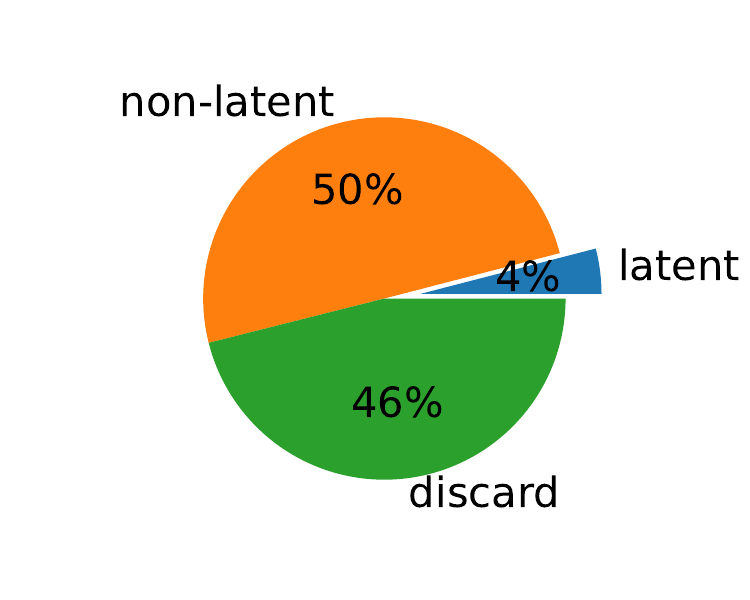}
    \vspace{-2.2em}
    \caption{CB}
    \label{fig:rq2_CB}
\end{subfigure}
%\hfill
\begin{subfigure}{0.18\textwidth}
    \includegraphics[width=\textwidth, trim = 0mm 5mm 0mm 10mm]{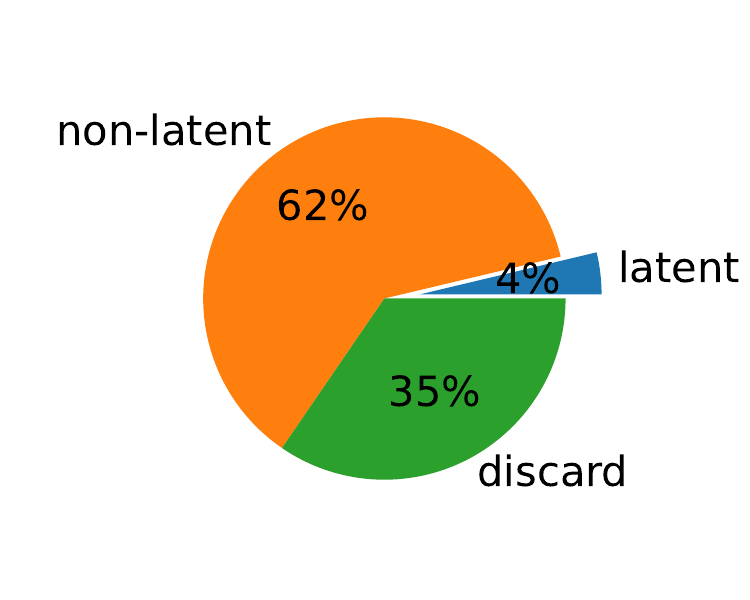}
        \vspace{-2.2em}
        \caption{VMC}
    \label{fig:rq2_VMC}
\end{subfigure}
%\hspace{-5em}
%\hfill
\begin{subfigure}{0.18\textwidth}
    \includegraphics[width=\textwidth, trim = 0mm 10mm 0mm 10mm]{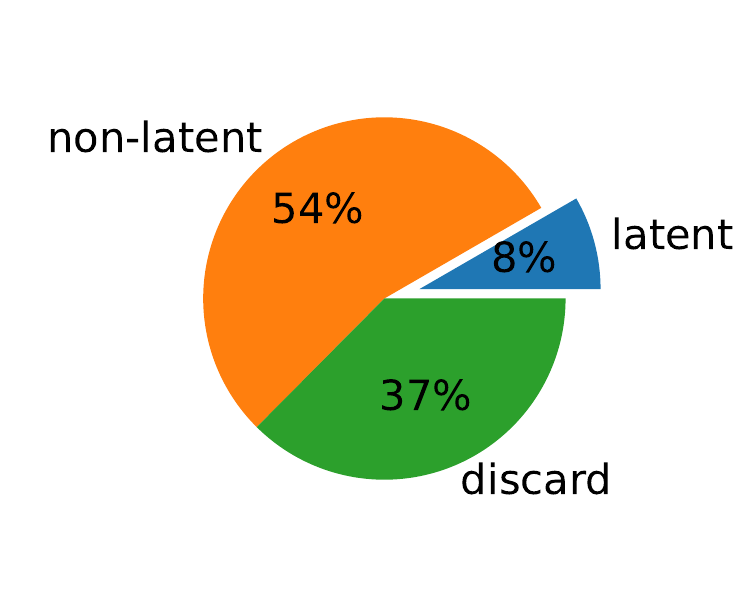}
        \vspace{-2.2em}
        \caption{MATH}
    \label{fig:rq2_MATH}
\end{subfigure}
%\hspace{-5em}
%\hfill
\begin{subfigure}{0.18\textwidth}
    \includegraphics[width=\textwidth, trim = 0mm 10mm 0mm 10mm]{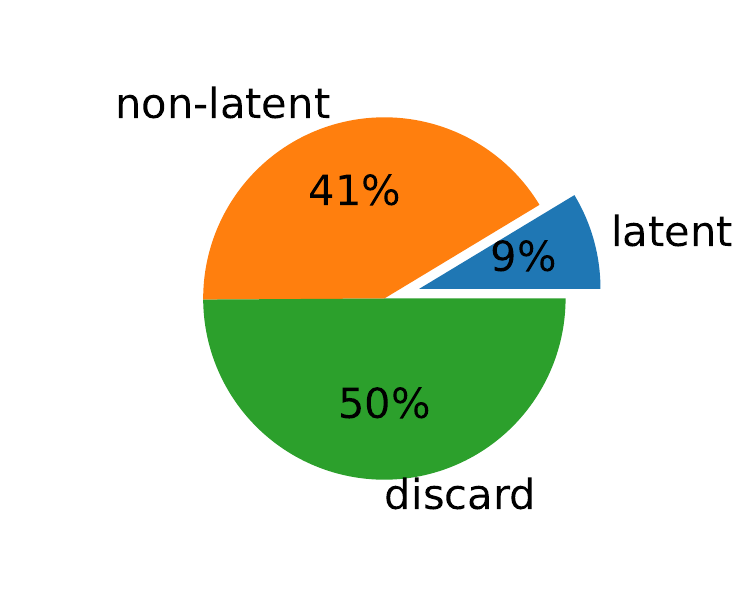}
        \vspace{-2.2em}
        \caption{NC}
    \label{fig:rq2_NC}
\end{subfigure}  
%\newline 

%\vfill 
\begin{subfigure}{0.18\textwidth}
    \includegraphics[width=\textwidth, trim = 0mm 10mm 0mm 10mm]{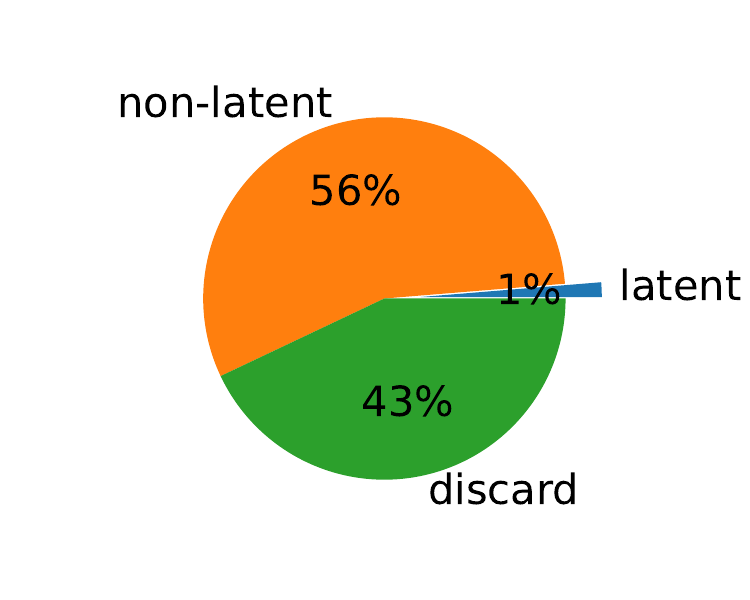}
        \vspace{-2.2em}
        \caption{NRET}
    \label{fig:rq2_NRET}
\end{subfigure}
\begin{subfigure}{0.18\textwidth}
    \includegraphics[width=\textwidth, trim = 0mm 10mm 0mm 10mm]{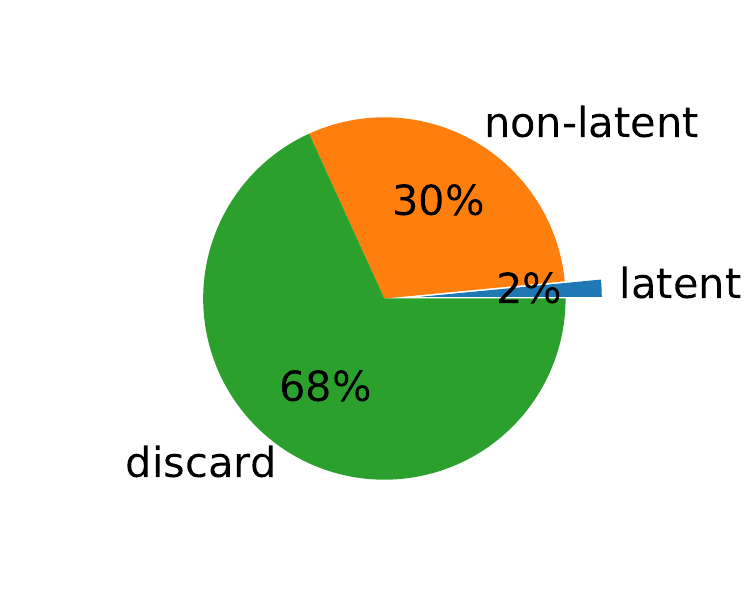}
        \vspace{-2.2em}
        \caption{INCR}
    \label{fig:rq2_INCR}
\end{subfigure}
\hspace{4mm}
\begin{subfigure}{0.18\textwidth}
    \includegraphics[width=\textwidth, trim = 0mm 10mm 0mm 10mm]{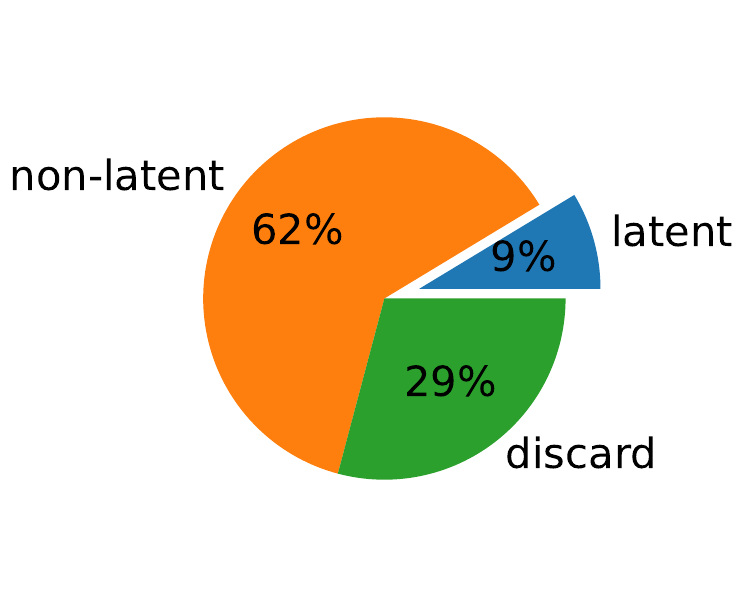}
        \vspace{-2.2em}
        \caption{ERET}
    \label{fig:rq2_ERET}
\end{subfigure}
\hspace{4mm}
\begin{subfigure}{0.18\textwidth}
    \includegraphics[width=\textwidth, trim = 0mm 10mm 0mm 10mm]{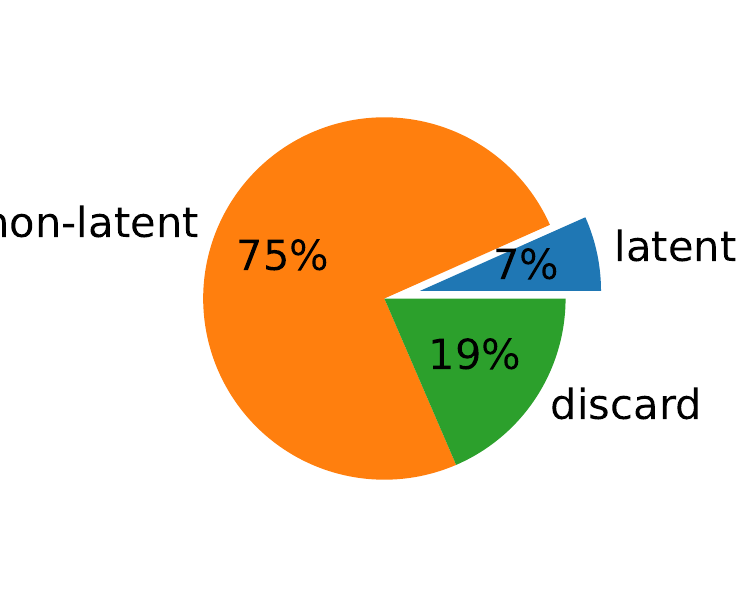}
        \vspace{-2.2em}
        \caption{BFRET}
    \label{fig:rq2_BFRET}
\end{subfigure}
    \vspace{-1.0em}
\caption{
%\todo{fix table arrangement} 
%Mutators and Mutant status. 
%Details of 
Prevalence of latent mutants,  non-latent mutants and discarded mutants for each  
mutation operator (aka mutator).  
%propeties 
%leading to Killed Mutants at mutant injection time. 
CB, VMC and MATH are among the most useful mutators for initial live mutants, whereas NC, NRET, INCR, ERET, and BRET are those useful to generate killed mutants. 
}
\label{fig:rq2_mut_op_pie}
    \vspace{-1.0em}
\end{figure}

\vspace{0.5mm}
\noindent
\textbf{Latent Mutants vs.  Non-latent Mutants vs.  Discarded Mutants:}  
Overall,  we observed that \textit{certain mutators are effective in producing latent mutants: 
%\cref{fig:rq2_mut_op_pie} shows that e
Eight (8) to nine (9) percent of the mutants generated by NC,  ERET,  and MATH mutators lead to latent mutants} (\textit{see} \cref{fig:rq2_mut_op_pie}).
This results suggests that some mutators 
%(e.g., NC,  ERET,  and MATH mutators) 
are more suitable for producing latent mutants.  
\cref{fig:rq2_mut_op_pie} further demonstrates that NC,  ERET,  and MATH mutators produce the most latent mutants and NRET produced the least latent mutants.  
%We believe this is because \todo{... why? intuitively}. 
%In addition,  w
We also observed that some mutators are more likely to lead to non-latent mutants and discarded mutants,  than other mutators.  \cref{fig:rq2_mut_op_pie} shows that INCR leads to the most (68\%) discarded mutants,  and BFRET leads to the least (19\%) discarded mutants.  
%Specifically,  68\% and 19\% of the mutants generated by the INCR mutator and BFRET mutator are discarded mutants,  respectively.  
Conversely,  INCR produces the least proportion (30\%) of non-latent mutants, and 
BFRET produces the most proportion (75\%) of non-latent mutants. 
Overall, this results imply that the choice of mutation operators is important in effectively producing latent mutants. 

\begin{result}
Mutator choice is relevant for the generation of latent mutants:
%since certain mutators are considerably effectively at producing latent mutants: 
%Up to o
One in eleven (9\% of) mutants generated by NC and ERET mutators lead to latent mutants. 
\end{result}

% lead to mutants that are not relevant when propagated.  
%While BFRET lead to only 19\% of discarded mutants.  
%We believe this is because \todo{... why? intuitively}.  
%Meanwhile,  the 
%VMC mutator is the most likely to produce non-latent mutants ( and 

%\revise{
%\fixme{}{include only five ops?} % Fig 3 might be too large and contain unimportant information as our focus is not on surviving and dead mutants. So, drop and include only the top five mostly used operators? 
%The majority of propagated mutants either survive to the end or become lost, i.e., dead, during the propagation, as shown in \cref{tab:tab_rq2_mut_status}. \cref{fig:rq2_mut_op_pie} compares the proportions of latent, \fixme{surviving}{}, and dead mutants for the eight mutators selected from the set of effective mutators for initial surviving (3) and killed mutants (5). The pie charts in the figure show that the usage of different mutators leads to different characteristics of mutants; for instance, VMC tends to generate more mutants that remain undetected to the end. 
%}

%\begin{result}
%XXXX
%\end{result}

\begin{table}[ht] 
\vspace{-0.5em}
\caption{
%Mutation Operators and Latent Mutants. 
Details of Mutation operators (aka mutators) 
leading to Latent Mutants in future project versions. 
JxPath, Collections and JacksonXml are excluded as they no longer have any mutant left.
}\label{tab:tab_rq2_rv_mut_mutop}
\vspace{-1.0em}
\scalebox{0.7}{
\begin{tabular}{l|rrrrrrrrrrr}
    \toprule
\textit{Perc (\%)} & MATH & CB & INCR & IN & NC & VMC & PRET & ERET & BFRET & BTRET & NRET \\
\midrule 
Lang & \cellcolor{green!25}7.14 & 3.42 & \cellcolor{green!25}16.67 & 0.0 & 6.35 & \cellcolor{blue!25}\textbf{17.39} & 0.0 & 5.88 & - & 0.0 & 0.0\\
Math & 24.03 & 9.46 & \cellcolor{green!25}25.0 & \cellcolor{blue!25}\textbf{46.67} & \cellcolor{green!25}34.48 & 19.14 & 11.32 & 0.0 & 0.0 & 6.25 & 0.0\\
Time & \cellcolor{green!25}0.0 & \cellcolor{green!25}0.0 & - & - & \cellcolor{green!25}13.33 & \cellcolor{blue!25}\textbf{23.08} & \cellcolor{green!25}0.0 & \cellcolor{green!25}0.0 & - & \cellcolor{green!25}0.0 & -\\
Closure & 0.0 & 0.8 & 0.0 & - & \cellcolor{blue!25}\textbf{11.23} & 0.41 & \cellcolor{green!25}2.33 & \cellcolor{green!25}4.0 & 1.85 & 1.08 & 0.46\\
Cli & \cellcolor{green!25}7.69 & \cellcolor{blue!25}\textbf{9.59} & 0.0 & - & 0.0 & \cellcolor{green!25}2.63 & 0.0 & 0.0 & 0.0 & 0.0 & 1.39\\
Compress & \cellcolor{green!25}4.0 & 0.62 & 0.0 & - & \cellcolor{green!25}4.0 & 2.55 & \cellcolor{blue!25}\textbf{4.65} & 0.0 & 0.0 & 0.0 & 0.0\\
Codec & 2.04 & \cellcolor{green!25}6.0 & - & - & \cellcolor{green!25}8.33 & 0.0 & - & 0.0 & 0.0 & \cellcolor{blue!25}\textbf{33.33} & 0.0\\
Csv & \cellcolor{green!25}0.0 & \cellcolor{green!25}0.0 & - & - & \cellcolor{green!25}0.0 & \cellcolor{blue!25}\textbf{9.09} & \cellcolor{green!25}0.0 & - & - & - & -\\
JacksonCore & 0.62 & 2.0 & 1.23 & - & \cellcolor{green!25}2.83 & 0.43 & \cellcolor{green!25}2.94 & \cellcolor{blue!25}\textbf{16.67} & 0.0 & 0.0 & 2.7\\
Jsoup & 2.9 & 3.72 & 0.0 & - & 6.9 & 3.47 & 0.0 & \cellcolor{blue!25}\textbf{17.39} & \cellcolor{green!25}16.0 & \cellcolor{green!25}10.64 & 1.27\\
\midrule
Total & \cellcolor{green!25}11.03 & 3.95 & 3.08 & \cellcolor{blue!25}\textbf{45.65} & \cellcolor{green!25}8.06 & 2.83 & 4.48 & 7.29 & 4.81 & 3.94 & 0.88\\
\bottomrule
TopFreq & \cellcolor{green!25}6 & 4 & 2 & 2 & \cellcolor{blue!25}\textbf{8} & 4 & \cellcolor{green!25}5 & 4 & 1 & 3 & 0\\
\bottomrule
\end{tabular}}
\vspace{-0.5em}
\end{table}

\begin{figure}[ht]
\vspace{-0.5em}
\centering
\begin{subfigure}{0.3\textwidth}
    \includegraphics[width=\textwidth, trim = 5mm 10mm 5mm 0mm]{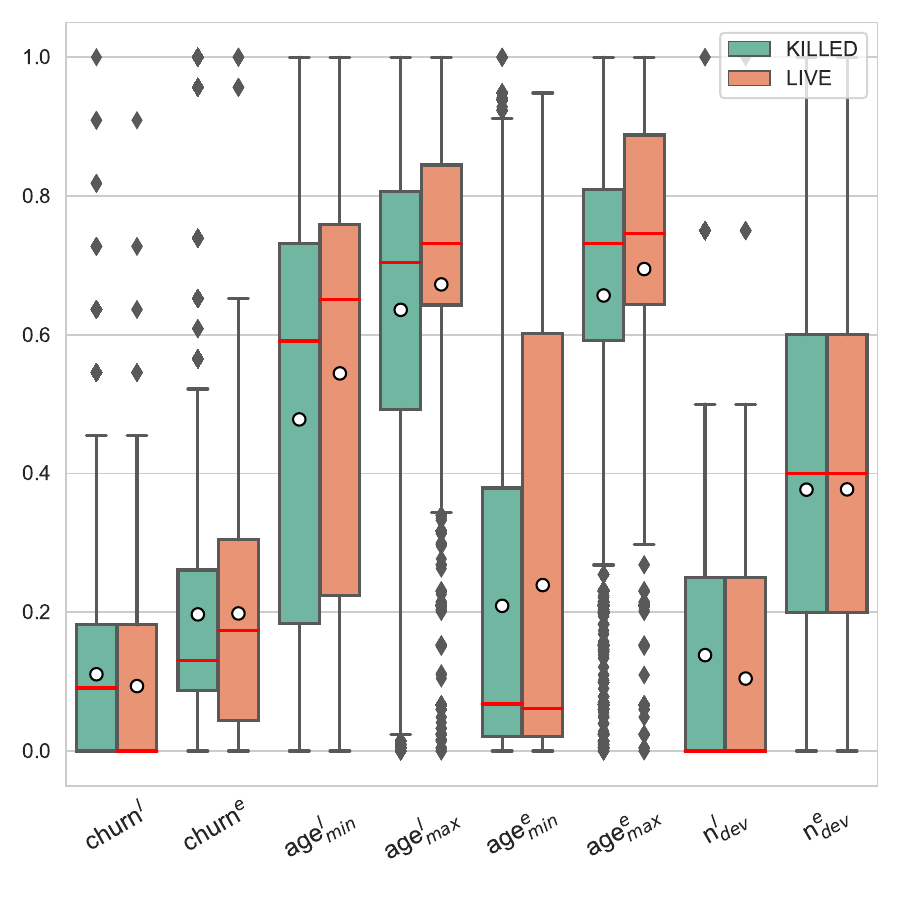}
    \caption{Lang}
    \label{fig:rq1_lang_chg}
\end{subfigure}
\hfill
%\begin{subfigure}{0.3\textwidth}
%    \includegraphics[width=\textwidth, trim = 20mm 10mm 14mm 0mm]{figure/rq1/Compress_chg.pdf}
%    \caption{Compress} % Or Josup
%    \label{fig:rq1_compress_chg}
%\end{subfigure}
%\hspace{-5em}
%\hfill
\begin{subfigure}{0.3\textwidth}
    \includegraphics[width=\textwidth, trim = 5mm 10mm 5mm 0mm]{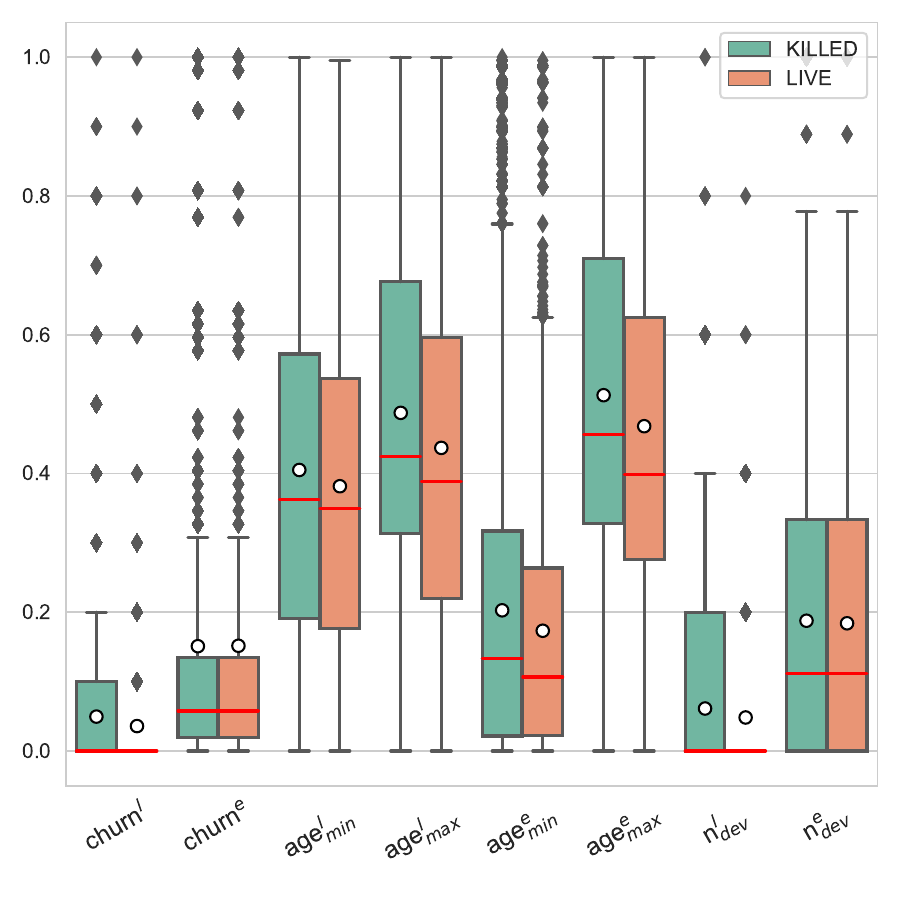}
    \caption{Closure}
    \label{fig:rq1_closure_chg}
\end{subfigure}
\hfill
\begin{subfigure}{0.3\textwidth}
    \includegraphics[width=\textwidth, trim = 5mm 10mm 5mm 0mm]{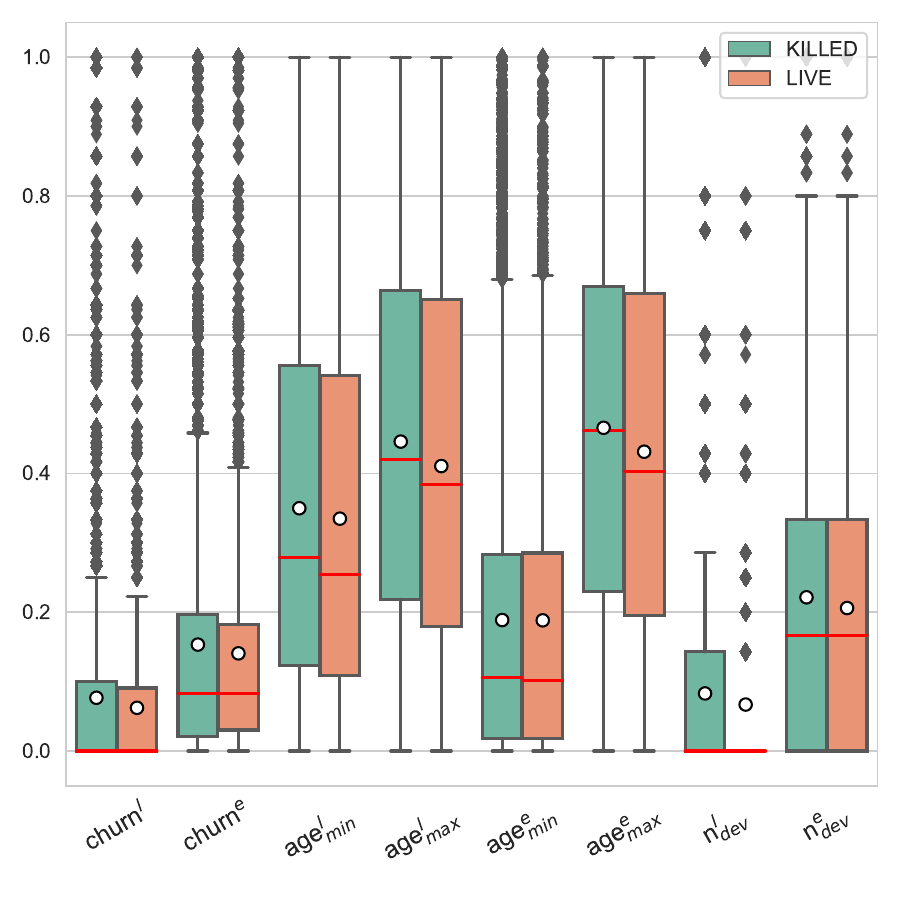}
    \caption{Total}
    \label{fig:rq1_total_chg}
\end{subfigure}
    \vspace{-1.0em}
\caption{
%Closure, Time, Math similar patterns: i.e., killed longer lived, Others: i.e., killed more recently changed  %(overall, churn more frequent -> common)
(White marks are the mean, and the red lines are the median. )
}\label{fig:rq1_chg}
    \vspace{-1.0em}
\end{figure}

\subsection{RQ3: Change Features}
\label{sub:rq3_change_features}
\begin{figure}[ht]
\vspace{-0.5em}
\centering
\begin{subfigure}{0.3\textwidth}
    \includegraphics[width=\textwidth, trim = 5mm 10mm 5mm 0mm]{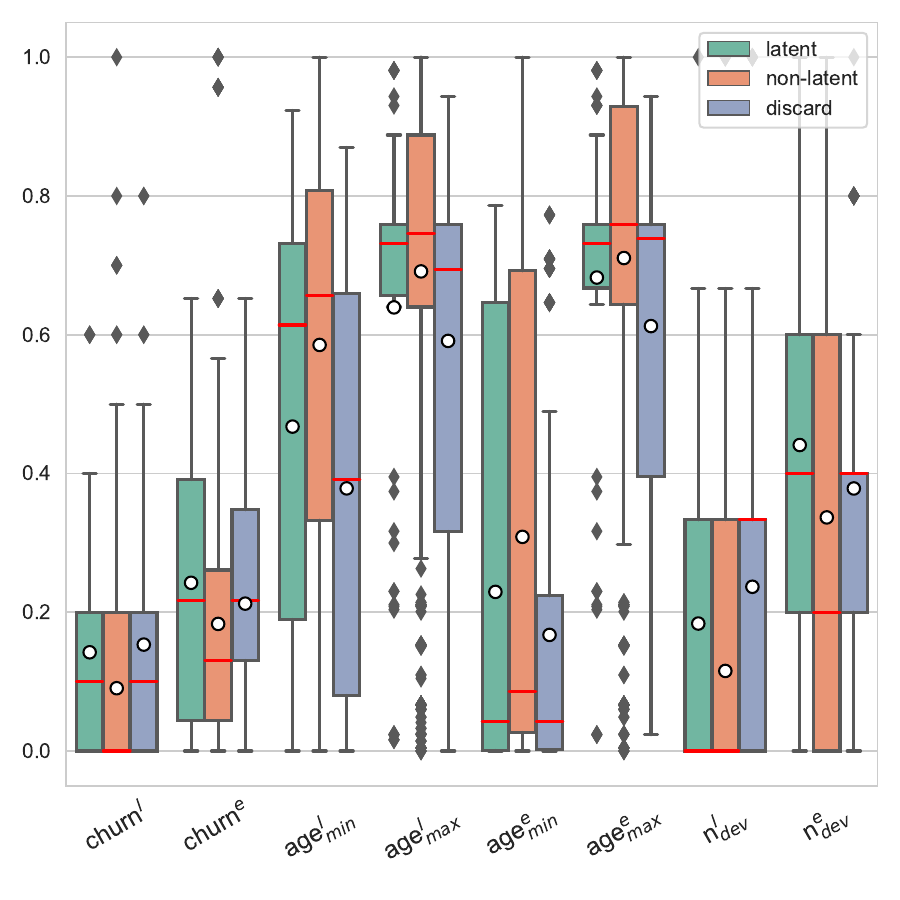}
    \caption{Lang}
    \label{fig:rq2_lang}
\end{subfigure}
\hfill
\begin{subfigure}{0.3\textwidth}
    \includegraphics[width=\textwidth, trim = 5mm 10mm 5mm 0mm]{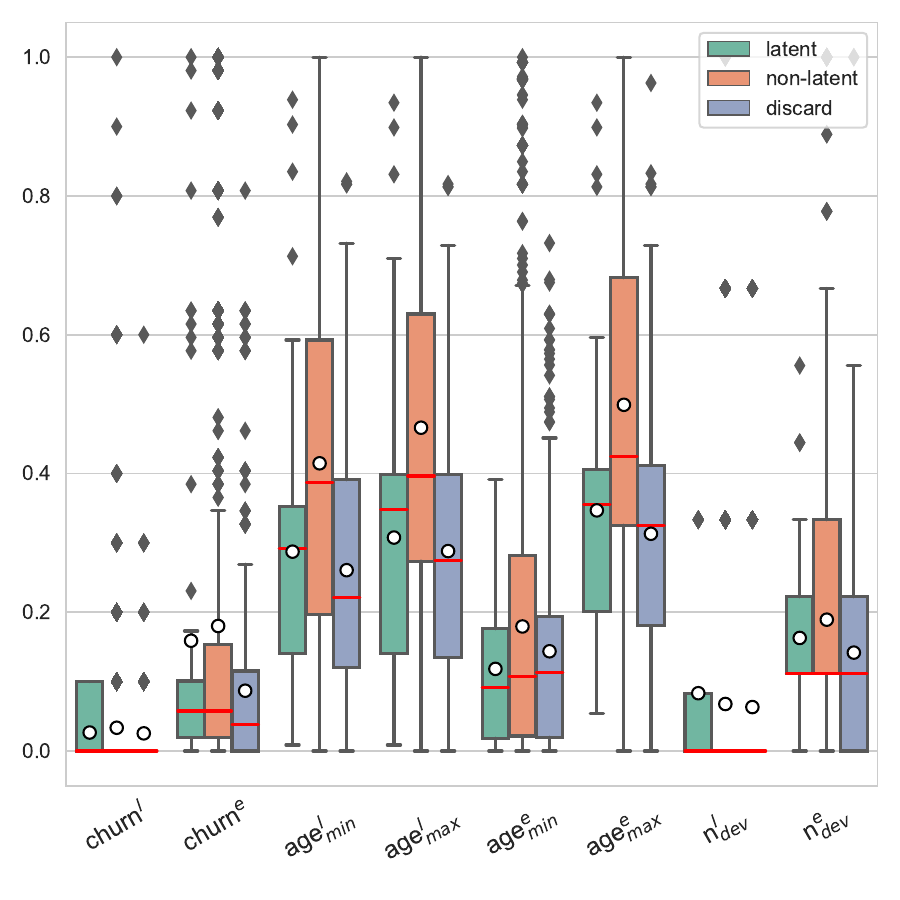}
    \caption{Closure}
    \label{fig:rq2_closure}
\end{subfigure}
\hfill
\begin{subfigure}{0.3\textwidth}
    \includegraphics[width=\textwidth, trim = 5mm 10mm 5mm 0mm]{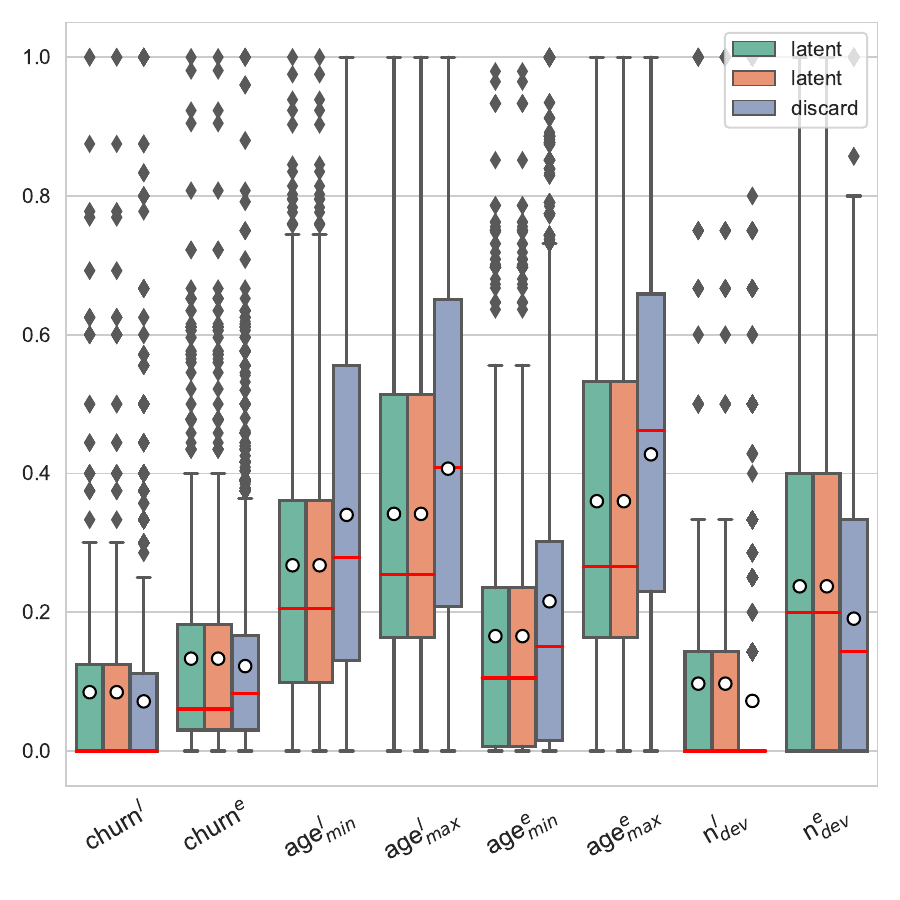}
    \caption{Total} % Or Josup
    \label{fig:rq2_total}
\end{subfigure}
    \vspace{-1.0em}
\caption{Trends in change features 
%for Latent
%,non-latent and discarded 
%mutants 
showing that the trend for latent mutants is strong in 
recently observed and relatively old but frequently changed program statements for most projects, except Lang.
(White marks are the mean, and the red lines are the median. )
%, the pattern of revealing line being  %(overall, churn more frequent -> common)
}\label{fig:rq2_chg}
    \vspace{-1.0em}
\end{figure}

This experiment first investigates how change features relate to the frequency of live mutants and killed mutants.   
\cref{sub:change_feature_collection} presents the historical change features employed in this experiment. 
\cref{fig:rq1_chg} illustrates the trend of the eight historical metrics between live mutants and killed mutants using two representative and all projects.  In addition, we examine how change features relate to the frequency of latent mutants, live mutants and  discarded mutants and \cref{fig:rq2_chg} illustrates our findings.  Finally,  we inspect the impact of change types on latent mutants (\textit{see}  \cref{tab:tab_rq2_revealed_mutant_cat}).  The first columns of each change category in \cref{tab:tab_rq2_revealed_mutant_cat} presents the \revise{number of latent ($L$), non-latent($N$), and discarded ($D$) mutants after the propagation. }
%results of these experiments.  
%\todo{figures and tables}

\vspace{0.5mm}
\noindent
\textbf{Live Mutants vs. Killed Mutants:}
We observed \textit{that killed mutants reside more often in the recently changed and introduced code, whereas the the opposite holds for live mutants} (\textit{see} \cref{fig:rq1_total_chg}).\footnote{We observe a similar trend across four different projects; some have mixed trend of the opposite depending on the metric.} In addition, mutants in the frequently changed statements tend to be slightly more detectable than those in rarely changed statements. 
These two observations imply that mutating actively maintained code results in more killed mutants.  We believe this is due to frequent testing by developers. However, we also often observe the opposite trend, as shown in \cref{fig:rq1_closure_chg} for Closure\footnote{Live mutants of around five projects often reside in the code that was recently changed or introduced compared to the killed mutants.} Here, mutating recently changed or introduced code leads to more live mutants. 
%\checkme{On inspection, we observed this is because the recent code has not yet been sufficiently tested, in contrast to the previous cases. }
\checkme{We conjecture that this may relate to the recent code being not sufficiently test, in contrast to the previous cases. }% I toned down this, b/c technically, I didn't check this, This is a suspcion 
%as opposite to the previously cases. 
We do not observe a clear trend for the number of developers, i.e., $n^l_{dev}$ and $n^e_{dev}$. 
Overall,  this result suggests that certain change features (e.g.,  recently changed, introduced or tested code) may be a good proxy for predicting killed mutants and live mutant.  
%Thus, we investigate the predict
%\todo{...} 
% the trend is less obvious. 
%Nevertheless, despite having different trends, all projects show differences in the past change tendencies of code with surviving mutants and killed mutants. 
%}

\begin{result}
Killed mutants are frequently found in under certain change conditions 
%are related to the 
(e.g.,  recently changed code) suggesting that change features may be a good proxy for predicting killed (or live) mutants. 
\end{result}

\vspace{0.5mm}
\noindent
\textbf{Latent Mutants vs Non-latent Mutants vs. Discarded Mutants:} \cref{fig:rq2_chg} shows that \textit{latent mutants (i.e., {\color{green}green}) often reside in the code relatively recently modified by developers and in fairly new lines or methods, compared to the mutants that have never been killed} (i.e., {\color{orange}orange}).  Overall,  the boxplots in \cref{fig:rq2_chg} show that these three mutant types (latent, live and discarded mutants) have different trends of past changes. Compared to the change trend we observed in \cref{fig:rq1_chg}, the projects that had different tendencies, such as Lang and Closure, now share a similar trend, mainly in the age of the mutated code ($age_{min}$ and $age_{max}$). However, we also observe the opposite, similar to the previous comparison between the original live and killed mutants. The trends in churn metric (i.e., change frequency) and the number of developers differ between mutant types and between projects.
\cref{fig:rq2_total} presents the combined results of mutants across projects. Despite individual projects having different characteristics, shared trends exist between mutants across projects.\footnote{Note that for discarded mutants, the trend is quite different from what has been found in individual projects. This is due to a certain project, in this case, Math, dominating the others by having a strong trend for discarded mutants.}

% tech-debt discussion

\begin{result}
Latent mutants are found in the code that the developer modified or introduced relatively recently, especially in comparison to non-latent mutants.
\end{result}

\begin{table}[ht] 
\vspace{-0.5em}
\caption{Detailed the relationship between mutants and different code changes types, and the lifespan. 
L/NL/D refer to the number of latent, non-latent, and discarded mutants under each change type (i.e., \textit{Semantic}, \textit{Refactoring}, \textit{NoChange}). 
Remaining columns report the number of mutants and the average life-span ($span$) of change-based, five mutant types (days)
}\label{tab:tab_rq2_revealed_mutant_cat}
\vspace{-1.0em}
\scalebox{0.65}{
\begin{tabular}{lr|r|rrrrrr|r|rrrrrr|r|rr}
\toprule
& & \multicolumn{7}{c|}{Semantic (and Refactoring (194))}  & \multicolumn{7}{c|}{Refactoring} & \multicolumn{3}{c}{NoChange} \\
& & \#$_{mut}$ & \multicolumn{2}{c}{SC$_C$} & \multicolumn{2}{c}{SC$_{NC}$} & \multicolumn{2}{c|}{total} & \#$_{mut}$ & \multicolumn{2}{c}{RC$_C$} & \multicolumn{2}{c}{RC$_{NC}$} & \multicolumn{2}{c|}{total} & \#$_{mut}$ &\multicolumn{2}{c}{NC$_{NC}$}\\
%proj & \#$^{all}_{rev}$ & \#$_{mut}$ & \#$^{latent}_{mut}$/\#$^{surv}_{mut}$/\#$^{dead}_{mut}$ & \# & dbt & \# & dbt & \# & span & \#$^{latent}_{mut}$/\#$^{surv}_{mut}$/\#$^{dead}_{mut}$ & \# & span & \# & span & \# & span & \#$^{latent}_{mut}$/\#$^{surv}_{mut}$/\#$^{dead}_{mut}$ & \# & span \\
proj & \#$_{rev}$ & L/NL/D & \# & span & \# & span & \# & span & L/NL/D & \# & span & \# & span & \# & span & L/NL/D & \# & span \\
\midrule
Lang & 27 & 1/44/25 & 1 & 1.5 & 0 & 0.0 & 1 & 1.5 & 3/3/4 & 0 & 0.0 & 0 & 0.0 & 3 & 2.0 & 35/571/147 & 35 & 111.6\\
Math & 79 & 15/171/30 & 1 & 16.2 & 14 & 68.0 & 15 & 64.5 & 0/0/0 & -- & -- & -- & -- & -- & -- & 121/660/445 & 121 & 85.6\\
Time & 10 & 0/2/0 & -- & -- & -- & -- & -- & -- & 0/0/0 & -- & -- & -- & -- & -- & -- & 5/71/1 & 5 & 245.1\\
Closure & 100 & 0/29/1 & -- & -- & -- & -- & -- & -- & 0/0/0 & -- & -- & -- & -- & -- & -- & 33/1639/540 & 33 & 144.4\\
Cli & 22 & 0/10/0 & -- & -- & -- & -- & -- & -- & 0/0/0 & -- & -- & -- & -- & -- & -- & 10/174/38 & 10 & 4.2\\
Compress & 43 & 2/41/43 & 1 & 56.9 & 1 & 319.9 & 2 & 188.4 & 0/0/1 & -- & -- & -- & -- & -- & -- & 24/503/475 & 24 & 147.1\\
Codec & 16 & 2/20/5 & 2 & 2.6 & 0 & 0.0 & 2 & 2.6 & 0/0/0 & -- & -- & -- & -- & -- & -- & 11/128/105 & 11 & 53.7\\
Collections & 2 & 0/0/0 & -- & -- & -- & -- & -- & -- & 0/0/0 & -- & -- & -- & -- & -- & -- & 0/4/0 & -- & -- \\
Csv & 11 & 0/9/9 & -- & -- & -- & -- & -- & -- & 0/0/0 & -- & -- & -- & -- & -- & -- & 1/106/18 & 1 & 0.8\\
JacksonCore & 21 & 1/0/22 & -- & -- & -- & -- & -- & -- & 2/0/2 & 0 & 0.0 & 1 & 228.4 & 3 & 232.7 & 48/330/2230 & 48 & 150.2\\
JacksonXml & 3 & 0/0/0 & -- & -- & -- & -- & -- & -- & 0/0/0 & -- & -- & -- & -- & -- & -- & 0/4/13 & -- & -- \\
JxPath & 9 & 0/2/0 & -- & -- & -- & -- & -- & -- & 0/0/0 & -- & -- & -- & -- & -- & -- & 0/120/13 & -- & -- \\
Jsoup & 79 & 4/16/22 & 3 & 0.4 & 0 & 0.0 & 3 & 0.4 & 0/0/0 & 0 & 0.0 & 1 & 238.1 & 1 & 238.1 & 46/399/774 & 46 & 87.1\\
\midrule
Total & 422 & 25/344/157 & 8 & 10.1 & 15 & 84.8 & 23 & 58.8 & 5/3/7 & 0 & 0.0 & 2 & 233.3 & 7 & 134.6 & 334/4709/4799 & 334 & 106.7\\
\bottomrule
\end{tabular}}
\vspace{-0.5em}
\end{table}

\vspace{0.5mm}
\noindent
\textbf{Impact of Change Types on Latent Mutants:} 
\checkme{Our experimental results show that \textit{latent mutants are often uncovered by changes in dependent code rather than changes to the mutants themselves: most (91.7\% of) the latent mutants we investigated did not undergo any changes until they are revealed.}}
\cref{tab:tab_rq2_revealed_mutant_cat} (\textit{NoChange} column) shows that most propagated mutants do not undergo changes that touch the program semantics or syntax;\footnote{The majority of changes are limited to style changes} this relates to the line granularity selected to check changes made to mutated code. Meanwhile, about half of these mutants live to the end, and developers overwrite the other half (discarded).  
The remaining 334 mutants are revealed in later commits. In fact, most of the latent mutants of all projects belong to this \textit{NoChange} -- 91.7\% by 334 out of 364 -- and, thereby, NC$_{NC}$ category, except Collections, JacksonXml, and JxPath, where none are revealing; \checkme{this suggests that the changes on the mutated line itself may not affect the revelation of mutants.} The \textit{Semantic} column of \cref{tab:tab_rq2_revealed_mutant_cat} further supports this suspicion. Despite containing only the mutants on the semantically changed code, the ratio of latent mutant (6.26\%) does not differ much from the ratio in NC$_{NC}$ (3.45\%). Only 15 mutants are purely refactored, and they are mainly from Lang. In our study, most of the mutants that underwent refactoring also underwent semantic changes: out of 524 mutants on the semantically changed code, 194 experienced code refactorings. These results imply that mutants are rarely revealed by code refactoring. 
%Overall, this result suggests that the changes that mutants go through may affect the final status of propagated mutants. 
%These observations imply that latent mutants are often uncovered by changes in dependent code rather than changes to the mutants themselves. 

\begin{result}
The type of code changes influences the revelation of latent mutants: Most (91.7\%) latent mutants are uncovered by changes in dependent code rather than changes to the mutants themselves. 
\end{result}

\begin{table}[ht] 
\vspace{-0.5em}
\caption{Lifespan (LS) and the number of revisions (\#$_{rev}$). % for Latent mutants, Non-latent mutants and Discarded mutants.  
Lifespan is in days. Each column has the value at \textit{25\%, 50\%, 75\%} and the average. 
Values within the bracket are computed without the studied period threshold.
}\label{tab:tab_rq4_lifespan}
\vspace{-1.0em}
\scalebox{0.65}{
\begin{tabular}{l|rrrr|rrrr|rrrr}
\toprule
& \multicolumn{4}{c|}{Latent} & \multicolumn{4}{c|}{Non-latent} & \multicolumn{4}{c}{Discard} \\
& 25\% &  50\% &  75\% & avg & 25\% &  50\% &  75\% & avg & 25\% &  50\% &  75\% & avg \\
\midrule
%LS (\# of days) & 1.97, 8.11, 130.43, 344.82 & 72.65 & 645.82, 931.29, 1358.26, 4050.24 & 1184.37 & 1.97, 160.09, 465.18, 3419.51 & 337.43\\
%\# of commits & 9.0, 30.0, 117.0, 857.0 & 103.77 & 628.0, 1062.0, 1739.0, 3965.0 & 1154.83 & 15.0, 108.0, 344.0, 2294.0 & 264.6\\
% below is when also restrict to 365 days for discarded mutants
%LS (\# of days) & 1.97, 8.11, 130.43, 344.82 & 72.65 & 645.82, 931.29, 1358.26, 4050.24 & 1184.37 & 0.0, 23.17, 155.33, 363.16 & 88.67\\
%\# of commits & 9.0, 30.0, 117.0, 857.0 & 103.77 & 628.0, 1062.0, 1739.0, 3965.0 & 1154.83 & 1.0, 46.0, 108.0, 920.0 & 106.87\\
%LS (\# of days)  & 8.1 & 51.4 & 210.5 & 344.8 & 103.5 & 635.2 & 927.4 & 1362.9 & 4050.2 & 1187.1 & 0.0 & 23.2 & 160.1 & 363.2 & 90.1\\
%\# of commits & 22.0 & 55.5 & 173.0 & 857.0 & 135.6 & 628.0 & 1045.0 & 1695.0 & 3965.0 & 1136.9 & 1.0 & 46.0 & 108.0 & 920.0 & 108.0\\
LS & 8 (30)& 51 (243)& 211 (642)& 104 (425) & 635 (382)& 927 (777)& 1363 (1153)& 1187 (949) & 0 (1)& 23 (161)& 160 (475)& 90 (341)\\
\#$_{rev}$ & 22 (47)& 56 (195)& 173 (509)& 136 (360) & 628 (329)& 1045 (824)& 1695 (1355)& 1137 (931) & 1 (16)& 46 (108)& 108 (344)& 108 (267)\\
\bottomrule
\end{tabular}}
\vspace{-0.5em}
\end{table}

\subsection{RQ4: Lifespan of Latent Mutants}
\label{sub:rq4_latent_mutant_lifespan}
This RQ examines the lifespan of latent mutants. \cref{tab:tab_rq2_revealed_mutant_cat} shows the lifespan of latent mutants underwent different changes, and \cref{tab:tab_rq4_lifespan} compares their lifespan to that of the non-latent and the discarded. 
\checkme{We found that \textit{latent mutants require 104 days on average to be caught by future tests after injection among those revealed within 365 days.} Discarded mutants have a similar but slightly smaller life span than latent mutants (90 days in \cref{tab:tab_rq4_lifespan}); non-latent mutants have up to 13.2 times the life span of latent or discarded mutants, as expected. For the number of commits within this lifespan, latent mutants went through an average of 136 revisions, while discarded and non-latent mutants went through 108 and 1137 revisions. 
%To sum up, these results suggest that latent mutants that we found are useful, allowing developers to address the fault linked to latent mutants, that can remain hidden throughout number of changes, beforehand. 
\checkme{To sum up, these results suggest that latent mutants indeed capture true technical debt in the code, demonstrating that these mutants can remain hidden throughout a number of changes.}
 %Discarded mutants have the shortest lifespan and underwent the smallest number of commits. %, implying that they may be located the most frequently changed r
\cref{tab:tab_rq4_lifespan} also contains the values without the threshold (within the bracket) to show that the findings do not depend on the studied period: i.e., the same trend is also observed in these values.
%\cref{tab:tab_rq4_lifespan} shows that discarded mutants have a similar but sligthly smaller life span than latent mutants (90.1 days). In cont However, non-latent mutants have up to 13.2 times the life span of latent or discarded mutants, as expected. This means that developers can catch the fault represented by latent mutants by having these latent mutants in advance. 
}

\cref{tab:tab_rq2_revealed_mutant_cat} shows that mutants that were semantically changed have smaller lifespan than those that have never gone through any changes (NC). However, when we compare mutants based on whether the change directly affects the revealing status, denoted by the subscript $_C$ and $_{NC}$, latent mutants that are indirectly caught by the changes on dependent code have similar life span regardless of whether the mutated code has been semantically changed or not: 84.8 days for SC$_{NC}$ and 106.7 days for NC$_{NC}$. 
Overall, these results emphasizes the potential benefits of predicting latent mutants.  

\begin{result}
Latent mutants exist for about 104 days before discovery, undergoing 136 revisions. This is similar to the lifespan of discarded mutants (90 days), but 11.4 times as low as the life span of non-latent mutants. %Overall, latent mutants capture true underlying technical debt u
\end{result}

\begin{table}[ht] 
\vspace{-0.6em}
\caption{Predictive performance of our Random Forest-based Latent Mutant Prediction Model. All values in cells are reported in average. $L$, $NL$ and $D$ refer to latent, non-latent, and discarded mutants. 
}\label{tab:tab_rq3_revealed_mutants}
\vspace{-1.0em}
\scalebox{0.7}{
\begin{tabular}{lr|rrr|rrr|rrr}
\toprule
& & \multicolumn{3}{c|}{Random Forest} & \multicolumn{3}{c|}{Random Forest (wo mutOp)} & \multicolumn{3}{c}{Random} \\
proj & \#$_{rev}$ & acc (L/NL/D) & bal$_{acc}$ & MAP & acc (L/NL/D) & bal$_{acc}$ & MAP & acc (L/NL/D) & bal$_{acc}$ & MAP\\
\midrule 
Lang & 27 & 0.89 (0.94/0.91/0.92) & 0.65 & 0.37 & 0.84 (0.94/0.87/0.87) & 0.61 & 0.37 & 0.34 (0.65/0.42/0.6) & 0.33 & 0.33 \\
Math & 79 & 0.82 (0.9/0.86/0.89) & 0.67 & 0.35 & 0.82 (0.89/0.85/0.89) & 0.63 & 0.31 & 0.34 (0.64/0.48/0.56) & 0.34 & 0.29 \\
Time & 10 & 0.89 (0.95/0.89/0.94) & 0.38 & 0.92 & 0.89 (0.95/0.89/0.95) & 0.37 & 0.92 & 0.35 (0.68/0.39/0.64) & 0.39 & 0.25 \\
Closure & 100 & 0.83 (0.99/0.83/0.84) & 0.63 & 0.45 & 0.77 (0.98/0.77/0.79) & 0.52 & 0.31 & 0.34 (0.66/0.42/0.59) & 0.33 & 0.15 \\
Cli & 22 & 0.86 (0.94/0.87/0.92) & 0.64 & 0.42 & 0.85 (0.94/0.85/0.91) & 0.62 & 0.39 & 0.34 (0.66/0.41/0.61) & 0.35 & 0.45 \\
Compress & 43 & 0.79 (0.97/0.8/0.8) & 0.59 & 0.3 & 0.78 (0.97/0.79/0.79) & 0.58 & 0.23 & 0.34 (0.66/0.5/0.51) & 0.34 & 0.25 \\
Codec & 16 & 0.89 (0.97/0.91/0.9) & 0.73 & 0.83 & 0.9 (0.97/0.91/0.91) & 0.76 & 0.81 & 0.33 (0.65/0.48/0.53) & 0.34 & 0.18 \\
Collections & 2 & 0.98 (1.0/0.98/0.98) & 0.98 & - & 1.0 (1.0/1.0/1.0) & 1.0 & - & 0.38 (0.7/0.38/0.68) & 0.38 & - \\
Csv & 11 & 0.94 (0.99/0.95/0.94) & 0.6 & - & 0.96 (0.99/0.97/0.96) & 0.62 & - & 0.33 (0.66/0.4/0.61) & 0.29 & - \\
JacksonCore & 21 & 0.96 (0.98/0.98/0.96) & 0.71 & 0.28 & 0.96 (0.98/0.98/0.97) & 0.73 & 0.33 & 0.34 (0.66/0.63/0.39) & 0.34 & 0.08 \\
JacksonXml & 3 & 0.66 (1.0/0.66/0.66) & 0.77 & - & 0.76 (1.0/0.76/0.76) & 0.84 & - & 0.36 (0.69/0.58/0.45) & 0.36 & - \\
JxPath & 9 & 0.85 (1.0/0.86/0.86) & 0.62 & - & 0.85 (1.0/0.85/0.85) & 0.6 & - & 0.34 (0.67/0.37/0.64) & 0.32 & - \\
Jsoup & 79 & 0.84 (0.95/0.86/0.86) & 0.61 & 0.29 & 0.84 (0.95/0.87/0.87) & 0.6 & 0.26 & 0.34 (0.65/0.56/0.46) & 0.34 & 0.28 \\
\midrule
Total & 422 & 0.87 (0.96/0.88/0.89) & 0.67 & 0.34 & 0.85 (0.96/0.87/0.87) & 0.64 & 0.32 & 0.34 (0.66/0.51/0.51) & 0.34 & 0.13 \\
\bottomrule
\end{tabular}}
\vspace{-0.8em}
\end{table}

%\begin{figure}[ht]
%\centering
%\begin{subfigure}{0.4\textwidth}
%    \includegraphics[width=\textwidth, trim = 5mm 10mm 5mm 0mm]{figure/pred_eval.pdf}
%    \caption{Balanced Accuracy}
%\end{subfigure}
%\begin{subfigure}{0.4\textwidth}
%    \includegraphics[width=\textwidth, trim = 5mm 10mm 5mm 0mm]{figure/pred_eval_map.pdf}
%    \caption{Mean Average Precision}
%\end{subfigure}
%\caption{
%}\label{fig:rq3_rf_pred}
%\end{figure}

\begin{table}[ht] 
\vspace{-0.6em}
\caption{Feature Importance for our Random Forest Latent Mutant Prediction Model.  The results are reported per fold (five fold). The cell highlighted in {\color{blue}blue} contains feature with the highest feature importance values. Those within the top three are highlighted in {\color{green}green}. 
%The last row contains the results of Spearman correlation analysis between the debt time and the feature. 
}\label{tab:tab3_fimp}
\vspace{-1.0em}
\scalebox{0.7}{
\begin{tabular}{l|rrrrrrrrr}
\toprule
fold & mutOp & l\_churn & l\_min\_age & l\_max\_age & l\_n\_authors & e\_churn & e\_min\_age & e\_max\_age & e\_n\_authors \\
\midrule
0 & 0.101 & 0.038 & \cellcolor{green!25}{0.161} & \cellcolor{green!25}{0.188} & 0.017 & 0.087 & 0.144 & \cellcolor{blue!25}{0.211} & 0.054\\
1 & 0.101 & 0.038 & \cellcolor{green!25}{0.162} & \cellcolor{green!25}{0.185} & 0.017 & 0.086 & 0.146 & \cellcolor{blue!25}{0.211} & 0.055\\
2 & 0.101 & 0.037 & \cellcolor{green!25}{0.159} & \cellcolor{green!25}{0.189} & 0.017 & 0.087 & 0.145 & \cellcolor{blue!25}{0.211} & 0.055\\
3 & 0.102 & 0.037 & \cellcolor{green!25}{0.162} & \cellcolor{green!25}{0.187} & 0.017 & 0.087 & 0.144 & \cellcolor{blue!25}{0.21} & 0.054\\
4 & 0.102 & 0.037 & \cellcolor{green!25}{0.16} & \cellcolor{green!25}{0.187} & 0.017 & 0.087 & 0.144 & \cellcolor{blue!25}{0.212} & 0.054\\
\bottomrule
\end{tabular}}
\vspace{-0.8em}
\end{table}

\subsection{RQ5: Latent Mutant Prediction}
\label{sub:rq5_latent_mutant_prediction}
Based on the findings in \textbf{RQ2} and \textbf{RQ3} (i.e.,  the relationship between latent mutants and change features or mutation operators), we employ the studied eight historical change features and mutation operators to train a classifier to predict latent mutants using a Random Forest Classifier.  We examine the predictability of latent mutants using the setup in \cref{sub:model_training} and the evaluation metrics defined in \cref{sub:model_performance}. \cref{tab:tab_rq3_revealed_mutants} presents the classification results of the final mutant status, i.e., \textit{latent}, \textit{non-latent}, and \textit{discarded}, using a Random Forest Classifier. We take the random prediction as our baseline to show that these latent mutants are identifiable. 

%\revise{
Overall, the RF classification model outperforms the random baseline, achieving an accuracy of 0.87 and a balanced accuracy of 0.67. For the accuracy per class within the parenthesis, in most cases, the model obtains an accuracy higher than 0.9 for the latent mutant class.\footnote{these class-wise values are often better than both accuracy and balanced accuracy, which takes the average of the sensitivity and specificity.}
While these accuracy values demonstrate that the classification can predict the final mutant status, they report the general prediction performance over all classes; as the latent mutant takes only an average of 3.5\% of the initial live mutants, these metrics may not be well-suited to evaluate the performance regarding the latent mutant identification. \checkme{Hence, we further adopt Mean Average Precision (MAP), evaluating how well latent mutants are ranked when sorting all live mutants in descending order of their likelihood to be latent.}
Overall, the RF model obtains higher MAP than the random baselines; for projects such as Time and Codec, the model acquires a MAP greater than 0.8. From these, we argue that latent mutants can be predicted using the historical features and the mutation operator. 
%}

%\revise{
\cref{tab:tab3_fimp} presents the feature importance of the trained models. Overall, the age features have higher feature importance values than the others. This conforms with the previous findings of different change trends for different mutant types. While historical features explain the characteristics of mutated code, mutation operators describe the mutation itself. As these two contain different information, we further investigate how the mutation operator affects the prediction by training another model without using mutation operator information (\textit{wo mutOp}). The middle column in \cref{tab:tab3_fimp} shows a slight yet constant decrease in the accuracy across projects. 
%}

\begin{result}
%\revise{
Latent mutants are identifiable using the historical features and mutation operators. Notably,  the historical features (age features) have the highest feature importance.
%}
\end{result}

\section{Threats to Validity} % (fold)
\label{sec:threats_to_validty}

\paragraph{Internal Validity} % (fold)
\label{par:internal_validity}

Threats to internal validity may relate to they way we propagate the mutants to future versions. We use GumTree~\cite{GumtreeASE14}, a popular syntax diff tool, to compare the changes between commits and identify where to inject mutants. By working at the AST node level, we reduce the risk of mutants being injected into wrong locations. For code refactorings, we employ RefactoringMiner~\cite{Tsantails_RMiner18icse}, a well-established refactoring mining technique studied in various research that needs to detect code refactoring. Another threat may relate to the flakiness of testing, which may result in mutants being accidentally killed. While the chance of latent mutants belonging to this case may not be high, we plan to investigate the impact of flaky tests in future work. 

% paragraph internal_validity (end)

\paragraph{External Validity} % (fold)
\label{par:external_validity}

Threats to external validity may limit our findings on latent mutants to Java projects. However, the characteristics of latent mutants that we investigated are mostly related to the evolution of code made by developers. Nonetheless, we intend to extend our subject pool to include projects written in other programming languages, such as C or C++.  
The limited usage of Pitest may pose another threat to the generalizability of our findings. In the experiment, we use the DEFAULTS group of mutators, which includes 11 basic mutation operators. Thus, the studied mutants may not represent the entire mutant set; using more sophisticated mutators may result in more interesting findings. Still, our results support the existence of latent mutants and their usefulness. We plan to extend our set of mutators to involve remaining operators in Pitest and recent language model-based techniques~\cite{tufano_icsme_2019,Degiovanni_ubert_ICSTW222}.

% paragraph external_validity (end)

\paragraph{Construct Validity} % (fold)
\label{par:construct_validity}

The
%is refers to the 
threat to construct validity of this work regards the approximation of the number/proportion of mutant types within a 365 day window.  However,  latent mutants may be revealed beyond this time window and non-latent mutants may become latent in the near future.  To mitigate this threat,  \revise{ we have further investigated the lifespan of latent mutants beyond 365 days,  particularly, to as much as over 1000 days (\textit{see} \textbf{RQ4} and \cref{tab:tab_rq4_lifespan}).}
Another threat is that our identification of latent mutants 
%mutant types 
is dependent on the use of a finite set of test suites and mutators. 
% employed in our experiments.  
We mitigated these threats by employing a large set of mutants and real world test suites.  We note that the employed test suites and projects are based on real software artifacts curated by Defects4J. Finally, other confounding factors include the equivalence and subsumption relations among the studied mutants types.  We mitigate the threat of mutant equivalence by using the standard methods in mutation testing to reduce the probability of generating equivalent mutants, e.g., we use PIT to ensure no common language frameworks are mutated.  In the future, we plan to study the subsumption relationship among latent mutants and the other mutant types in this work.

%of 

%\todo{what to write...}
%One potential threat to the construct validity is the accuracy of matching between code elements. We use GumTree, a well-studied syntax-aware diff tool, to map AST nodes directly and indirectly through extended APIs of RefactoringMinder. RefactoringMinder supports mapping between more than one element. This support of various mappings can affect the propagation of mutants to following commits since the mutant can spread to multiple places. We thereby take into account the potential many-to-many, one-to-many or many-to-one mappings between code elements (i.e., AST nodes) while propagating the mutants to the subsequent versions. However, none of those mutated elements underwent such refactoring, and thus, none propagated to multiple places during the investigation. 

% dependent line definition: limited to the same file ->future work 

% paragraph construct_validity (end)

% section threats_to_validty (end)
\section{Related Work} % (fold)
\label{sec:related_work}

Mutation testing has been studied as one of the most effective testing criteria for its ability to reveal real faults~\cite{ChekamPTH17}. However, the high cost involved in mutation testing hinders its adoption in practice~\cite{Ammann_2014}. Various approaches have been proposed to address this issue, mainly aiming at selecting specific mutant types~\cite{OffuttLRUZ96}. This is typically happening at random ~\cite{HintTestDataSelection1978} or based on the characteristics of mutated location using static control flow graphs~\cite{SunXLZ17}. 

The objective of  mutant selection is to identify those mutants that can effectively assist testing, for instance, the mutants that can guide testing to where and what to test. Hence, existing approaches attempt to define and evaluate the \textit{interestingness} of mutants, selecting the mutants based on the measured values~\cite{PetrovicI18,JustKA17,Mirshokraie0P15,titcheu2020selecting}. Petrovic and Ivankovic~\cite{PetrovicI18} used arid mutants in the code AST to filter out the mutants that are likely to produce unproductive mutants. Just et al. leveraged AST parent and child nodes to select high-utility mutants~\cite{JustKA17}. Mirshokraie et al. \cite{Mirshokraie0P15} regard the mutants as useful when they are killable, using complexity and test execution information to differentiate them. Titcheu et al. consider mutants to be important when they are coupled with real-faults, identifying such mutants with static features, such as the data flow and complexity~\cite{titcheu2020selecting}. 

Recent work proposed to learn the characteristics of useful mutants~\cite{tufano_icsme_2019,GargODCPT23,Degiovanni_ubert_ICSTW222}. Tufano et al. \cite{tufano_icsme_2019} proposed to use Neural Machine Translation to learn how to mutate from bug-fixes. Garg et al. also employed Neural Machine Translation but with a different aim of identifying subsuming mutants among the existing mutants rather than generating them~\cite{GargODCPT23}. Degiovanni et al. used a pre-trained language model to generate natural mutants that resemble the developer's code~\cite{Degiovanni_ubert_ICSTW222}. 

Our study differs from these existing studies by approaching the problem from a different perspective. Instead of inspecting the links of mutants to the current test suite, we investigate their potential link to future tests. Specifically, we define and identify latent mutants coupled with latent faults that are hidden for now but later revealed by future tests. We further inspect whether these latent mutants are predictable by exploring their characteristics concerning software evolution. Hence, the findings of our study can complement existing work, extending the usefulness of mutants to future testing.

\section{Conclusion} % (fold)
\label{sec:conclusion}
In this paper we studied the co-evolution of mutants along software changes. Based on this analysis we identified new classes of mutants that can help developer improve their test suites. In particular we identified \textit{latent} mutants, i.e., the mutants that remain live in our current version but are killed in a later one. We then showed that we can predict these mutants using change-related features and thus, allowing developers to target them in advance. Overall, our results show that latent mutants are less than 1\% of all mutants, help improving test suites (as they lead to effective tests) and can last in time, thereby providing value that can stand in time.   

\bibliographystyle{ACM-Reference-Format}
\bibliography{newref}

\end{document}